\documentclass[fleqn,usenatbib,useAMS]{mnras}

\usepackage{graphicx}	
\usepackage{amsmath}	
\usepackage{amssymb}	
\usepackage{multicol}        
\usepackage{bm}		
\usepackage{pdflscape}	

\usepackage[T1]{fontenc}
\usepackage{ae,aecompl}

\usepackage{newtxtext,newtxmath}

\title[SMC Star Formation events]{Discrete star formation events in the central Bar of the Small Magellanic Cloud}

\author[A. Strantzalis et al.]{
A. Strantzalis$^{1}$\thanks{E-mail: a_strantzalis@yahoo.gr}, D. Hatzidimitriou$^{1,2}$, A. Zezas$^{3,4,5}$, V. Antoniou$^{5,6}$, S. Lianou $^{2}$
\newauthor and S. Tsilia $^{1}$
\\
\\
$^{1}$Department of Physics, National and Kapodistrian University of Athens, Panepistimiopolis, Zografos, GR15784, Greece \\
$^{2}$IAASARS, National Observatory of Athens, Vas. Pavlou and I. Metaxa, 15236 Penteli, Greece\\
$^{3}$University of Crete, Physics Department \& Institute of Theoretical \& Computational Physics, 71003 Heraklion, Crete, Greece\\
$^{4}$Foundation for Research and Technology-Hellas, 71110 Heraklion, Crete, Greece\\
$^{5}$Department of Physics, Box 41051, Science Building, Texas Tech University, Lubbock, TX 79409-1051, USA\\
$^{6}$Harvard-Smithsosian Center for Astrophysics, 60 Garden Street, Cambridge, MA 02138, USA\\
}

\date{Accepted 2019 September 03. Received 2019 September 03; in original form 2019 June 25}

\pubyear{2019}

\begin{document}
\label{firstpage}
\pagerange{\pageref{firstpage}--\pageref{lastpage}}
\maketitle

\begin{abstract}
We present the results of the photometric analysis of a large part of the main body of the Small Magellanic Cloud. Using the 6.5m Magellan Telescope at the Las Campanas Observatory in Chile, we have acquired deep B and I images in four fields (0.44 degree each in diameter), yielding accurate photometry for 1,068,893 stars down to 24$^{th}$ magnitude, with a spatial resolution of 0.20 arcsec per pixel. Colour-magnitude diagrams and (completeness corrected) luminosity functions have been constructed, yielding significant new results that indicate at least two discrete star formation events over a period from 2.7 to 4 Gyr ago. Also, we have derived star formation rates as a function of look back time and have found enhancements of SF between 4-6 Gyr and at younger ages. 


\end{abstract}




\section{Introduction}

At a distance of about 62kpc ($61\pm1 kpc$, \citealt{Hilditch2005}; $61 \pm 3 kpc$, \citealt{Inno2013}; $61.94\pm0.57 kpc$, \citealt{deGrijs2015}, $62.0\pm0.3 kpc$, \citealt{Scowcroft2016}), the Small Magellanic Cloud (SMC) is the second nearest dwarf irregular galaxy to our own, the closest being the disrupted Canis Major dwarf \citep{Martin2004, Conn2008}. The SMC interacts \citep[for a review]{D'Onghia2016} both with its neighbouring Large Magellanic Cloud (LMC) and with the Milky Way Galaxy (MW).  According to \citet{Besla2016} and \citet{Hammer2015}, the MCs are on their first infall towards the MW, having entered the Local Group as a small group of dwarf galaxies \citep{Sales2017}. Due to the relatively small mass of the SMC ($M_{dynamical} \simeq 2.4 \times 10^{9} M_{\odot}$, \citealt{DiTeodoro2019}) compared to the LMC and the MW,  its star and cluster formation histories, morphology and overall dynamics are expected to be significantly affected by these interactions.  Large scale features such as the Magellanic Bridge, the Magellanic Stream and the Leading Arm are attributed to these interactions (the first is caused by interactions between the Magellanic Clouds (MCs) and the latter two are due to interactions with the MW galaxy \citep{D'Onghia2016}). Moreover, the SMC is known to display complex geometry with significant depth at different lines of sight, associated with its disrupted morphology \citep[e.g.][]{Hatzidimitriou1993,  Subramanian2012, Scowcroft2016, Ripepi2017}.

\begin{figure*}
\centering
\includegraphics[height=9cm,width=15cm]{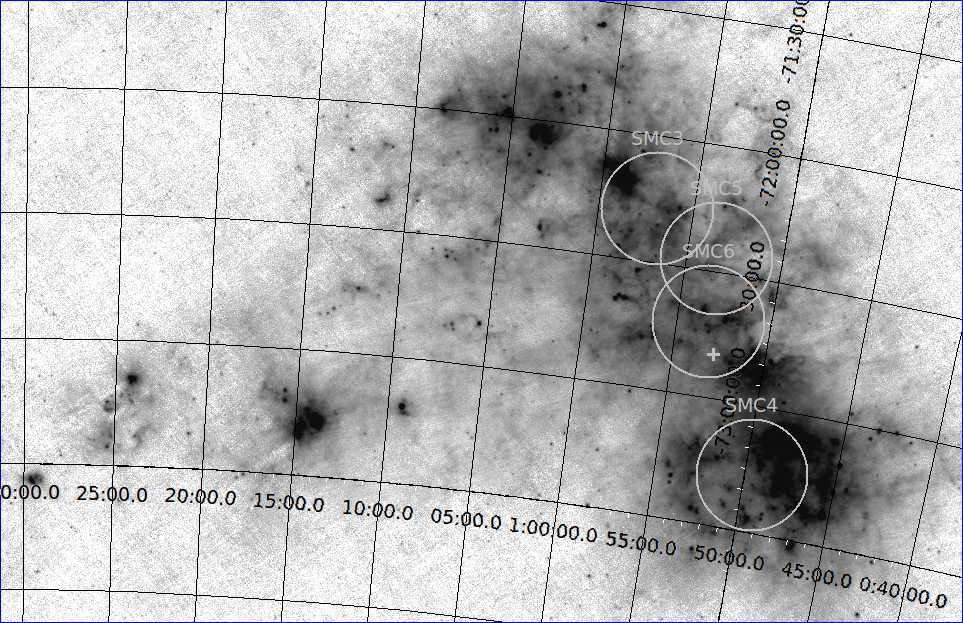}
\caption{The location of four fields studied in this work is marked on a Spitzer-MIPS 24$\mu$ map of the SMC \citep{Gordon2011}. The circles indicate the field of view of the IMACS data. 
\label{fig1}
}
\end{figure*} 

The stellar populations and star-formation history (SFH) of the SMC have been the subject of numerous investigations over the past 4 decades \citep[e.g.][]{Gardiner1992, Noel2007, Cignoni2013, Nidever2017, Rubele2018}, employing both ground-based and space observations. More recent studies are generally characterised by increased photometric accuracy and depth, improved spatial resolution and  increased sophistication of stellar models and SFH simulation codes. The first extensive study of the SFH of the SMC was performed by \citet{Harris2004}, who studied the central area (4 x 4.5 deg$^{2}$) of the galaxy, using data from the Magellanic Cloud Photometric Survey obtained with the 1-m Swope Telescope \citep{Zaritsky2002}. They concluded that approximately 50\% of the stars that have ever been formed in the SMC, did so more than about 8.4 Gyr ago, and that only a small amount of stars formed between 8.4 and 3 Gyr ago. They also claimed that during the last 3 Gyr there has been a rise in the mean star-formation rate (SFR) with three main bursts occurring at 2.5, 0.4 and 0.06 Gyr ago. 
The claim for low SF rate between 3 and 8 Gyr ago has been largely challenged by several subsequent studies, with the more recent one based on the Vista Magellanic Cloud Survey (VMC, \citealt{Rubele2018}). In the latter study it has been found that half of the stellar mass in the SMC formed prior to an age of 6.3 Gyr, while $\simeq$80\% of the stellar mass formed between 8 and 3.5 Gyr ago. Similar results, supporting enhanced star formation at intermediate ages between $\simeq$4 and $\simeq$6 Gyr have also  been found in specific (small) regions (\citealt{Noel2009,Cignoni2012,Weisz2013}).
 Concerning the oldest population, as exemplified by RR-Lyrae variables (e.g. \citealt{Muraveva2018}) and one bona-fide old star cluster, NGC 121 (\citealt{Glatt2008}), it comprises only a small percentage of the total stellar mass of the galaxy \citep{Soszynski2002}. \citet{Noel2009} and \citet{Sabbi2009} provide an age of about 12Gyr for the oldest stars although they both agree that SF was low at that time.

 Over the past $\sim$500Myr there has been a significant enhancement in SF,  strongly concentrated in the Bar and the Wing areas \citep{Cignoni2013,Sabbi2009}. Stars younger than $\sim$100Myr have a highly inhomogeneous spatial distribution \citep{Sabbi2009}, with a clear offset to the North-East and the Wing regions.

 Additional tracers of the SFH of the SMC include variable stars and AGB stars, although these involve a smaller number of objects. \citet{Rezaeikh2014} use long period variable stars to find two formation epochs, one at $\sim$6 Gyr and another at 0.7 Gyr. \citet{Cioni2006} use AGB stars to suggest that old stars (7-9Gyr) are located at the periphery of the SMC, while younger stars (< 7Gyr) are present towards the direction of the LMC.

The current study provides an extended as well as high spatial resolution survey of the central regions of the SMC, which allows to search for direct evidence  of distinct periods of enhanced SF  in the densest regions of the SMC Bar. We have obtained $B$ and $I$ images  with the 6.5m Magellan Telescope, using adaptive optics, resulting in high spatial resolution.  In Section 2 we present the observational data, in Section 3 we describe the data reduction and the estimation of the completeness of our data using artificial star experiments, in Section 4 we present the results, discuss them and  compare them with previous studies, and  in Section 5 we present the conclusions of our study.

\section{Observations}

Observations of four fields in the SMC were obtained with the 6.5m Magellan Telescope at the Las Campanas Observatory in Chile on October 4th, 2004, using the Inamori Magellan Areal Camera and Spectrograph (IMACS) \citep{Dressler2011}. IMACS is a wide-field imager and multi-object spectrograph with an eight CCD mosaic. In its f/2 configuration it provides a 0.44 deg diameter field of view with 0.2''  pixel size. The images were taken through B and I standard Johnson-Cousins filters. Each exposure was 120s long in $B$ and 30s in $I$. Each image was constructed from a total of 6 dithered single exposures in $B$ and 2 in $I$. Table 1 shows the log of the observations. The first column gives the ID of the field, columns 2 and 3  the coordinates of the center of each field, column 4 the filter through which the observations were performed, column 5 the total exposure time and column 6 the average airmass. 
{Fig.~\ref{fig1}} shows the location of the four  fields observed, overlayed on an infrared image of the SMC obtained with Spitzer (MIPS, 24 $\mu$). Each field covers 0.14 $\rm{deg}^2$ on the sky. The total area covered by the four fields is 0.5 $\rm{deg}^2$. The overlap regions between fields SMC3 and SMC5 and fields SMC5 and SMC6 were used to assess possible systematic errors in the photometry.

\twocolumn[
  \begin{@twocolumnfalse} 
  \maketitle{Table 1: Log of Observations}
	\begin{center}
    
			\begin{tabular}{ c c c c c c c c c c }
				\hline
				Field & RA (J2000) & Dec(J2000)& Filter &  Exposure Time (s)& Airmass \\
				SMC3 & 0:56:53.5 & -72:17:07.9 &  B&  6x120 & 1.377 &  \\
				SMC3 & 0:56:53:1 & -72:17:08.0 &  I&  2x30  & 1.375 &  \\
				SMC4 & 0:49:35.0 & -73:16:23.5 &  B&  6x120 & 1.628 &  \\
				SMC4 & 0:49:29.6 & -73:16:07.4 &  I&  2x30  & 1.632 &  \\
				SMC5 & 0:53:22.1 & -72:26:31.7 &  B&  6x120 & 1.421 &  \\
				SMC5 & 0:53:22.4 & -72:26:30.1 &  I&  2x30  & 1.421 &  \\
				SMC6 & 0:53:15.6 & -72:41:57.1 &  B&  6x120 & 1.712 &  \\
				SMC6 & 0:53:09.9 & -72:41:57.4 &  I&  2x30  & 1.711 &  \\ 
				\hline
			\end{tabular}
	\end{center}
  \end{@twocolumnfalse}
]

\section{Data Reduction}

\subsection{Photometry}

The images were bias subtracted and flat fielded using the IRAF/CCDPROC tool \citet{Tody1993}. Astrometry was achieved using the 2MASS \citep{Skrutskie2006} catalog as reference. The absolute astrometric accuracy of the reference catalog is approximately 0.1 arcsec. Accepted astrometric solutions had positional scatter less than 0.3 arcsec. The final mosaic image in each filter and field  was constructed using SWarp, which re-samples and co-adds FITS images using any arbitrary astrometric projection defined in the WCS standard \citep{Bertin2002}. The indivindual frames were median combined to produce a ''TAN'' projected image which is also corrected for distortions in the individual CCD images.

Photometry was performed using the DAOPHOT package  \citep{Stetson1987} in IRAF, separately for each constituent CCD. This choice was dictated by the positional variations of the point spread function (PSF) and of the background. The PSF was determined from 150 bright unsaturated stars  using PSTSELECT.  PSF photometry was then performed for all sources detected in each CCD mosaic using ALLSTAR by applying the corresponding PSF. The detection threshold was set at 4$\sigma$ above the background. The recovery of faint and/or partially blended sources was achieved by repeating the procedure with the same PSF (after removing the already measured sources) four times. The measurement of the magnitudes was performed on the initial image using the complete list of the detected stars. 

\begin{figure*}
	\centering
	\includegraphics[height=5cm,width=8cm]{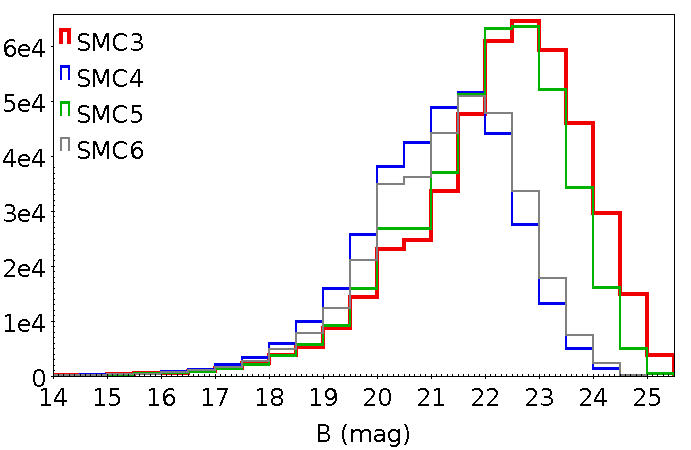}
	\includegraphics[height=5cm,width=8cm]{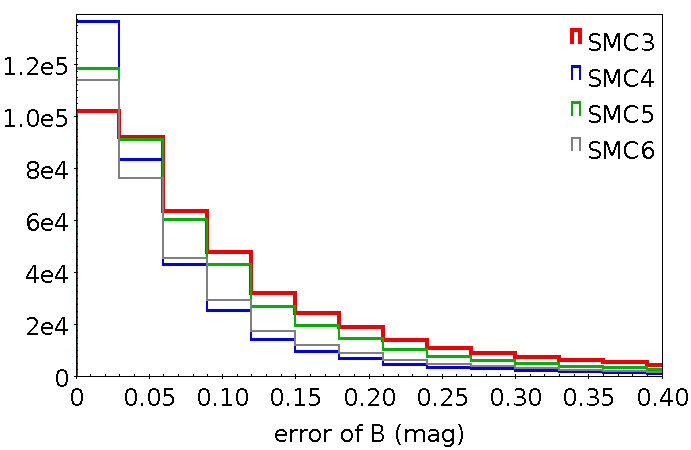}
	\includegraphics[height=5cm,width=8cm]{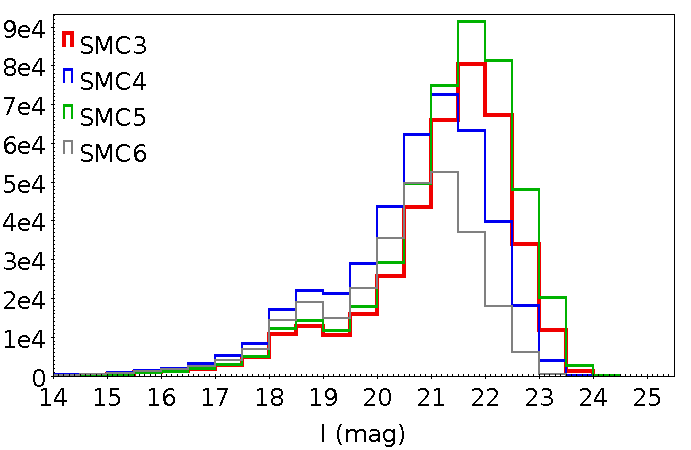}
	\includegraphics[height=5cm,width=8cm]{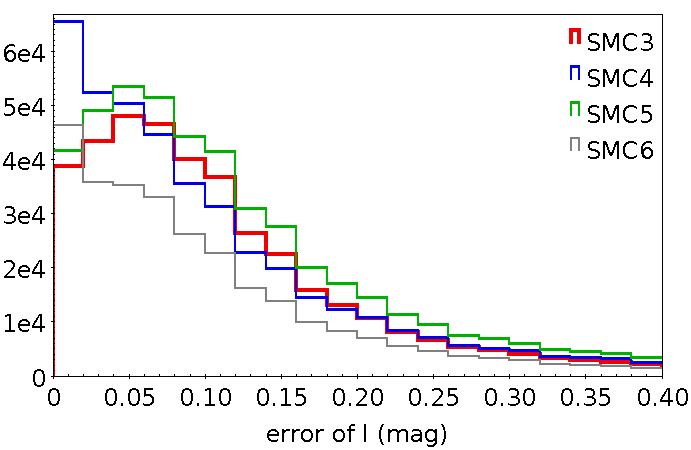}
	\caption{Histograms of the  calibrated magnitudes (left panels) and the distribution of errors (right panels) for the two filters ($B$ in the upper panels and $I$ in the lower panels) of all sources detected in each of the four fields under study. The red line corresponds to SMC3, the blue line to SMC4 the green line to SMC5 and the grey line to SMC6. }
\label{fig2}
\end{figure*}

\begin{figure*}
	\centering
	\includegraphics[height=5cm,width=8cm]{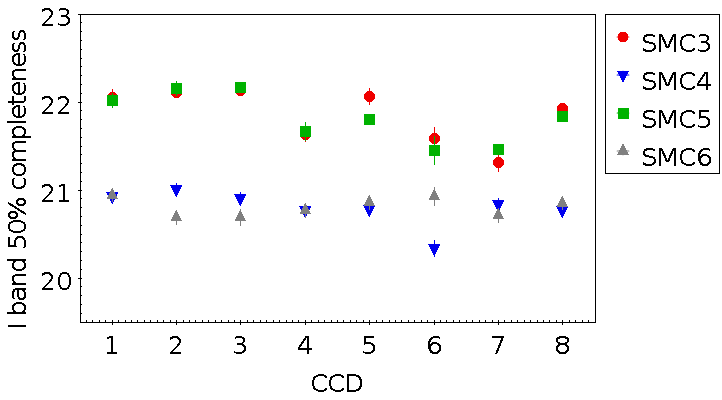}
	\includegraphics[height=5cm,width=8cm]{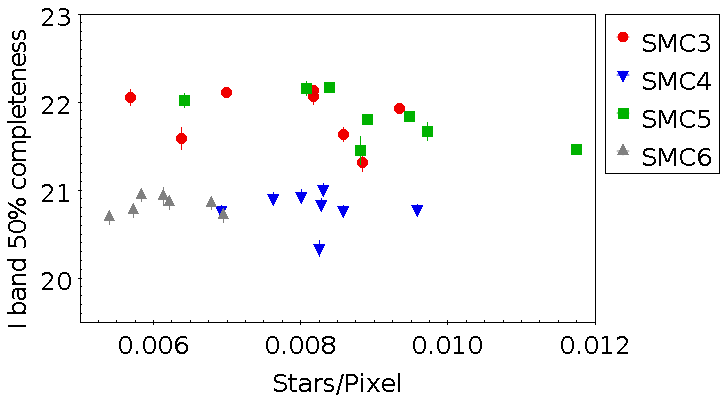}
	\caption{  Diagrams of the positional dependence of the level 50\% completeness within each CCD (left panel) and  the dependence of the level 50\% completeness values on stellar density (right panel), for the $I$ filter, for all fields.
	}
	\label{fig3}
\end{figure*}

Absolute photometric calibration was achieved by using a set of secondary standards selected from isolated relatively bright stars ($B<17.5$mag and $I<18$mag) in the \citet{Zaritsky2002} photometric catalog. It must be noted that a significant percentage of stars in the Zaritsky catalog were actually resolved into at least two sources in our data. Clearly, such cases were not included in the sample used for the photometric calibration. The calibration correction was performed separately for each CCD image and involved only a zero-point term, as the colour term was negligible. In Table 2 we report, for the different CCDs  the range of zero-point values  obtained (column 3) per field (column 1) and filter (column 2), as well as the total number of sources detected, N$_{tot}$ (column 4) and the number of sources for which the error in the instrumental magnitude was below 0.2mag, N$_{\sigma\leq0.2}$ (column 5). In most cases more than 85\% of the complete list of the sources fulfilled this criterion. Using the zero-point corrections for each CCD we derived the magnitude for each star. {Fig.~\ref{fig2}} shows the $B$ and $I$ magnitude distribution for sources in the four fields studied, as well as  the corresponding error distributions.  These errors include both instrumental magnitude errors  and  zero-point uncertainties. In the rest of the paper we use only stars with photometric instrumental errors below 0.2 mag. The saturation limit for all fields is $\simeq$ 14 mag. It is clear from these distributions that the observations of SMC3 and SMC5 are deeper in both filters. Photometry in the other two fields is affected by local and/or global background variations: SMC6 was affected by twilight, leading to relatively high sky values;  SMC4, as is immediately obvious by inspection of Fig.~\ref{fig1},  suffers from high and spatially variable interstellar absorption which affects the depth (and accuracy) of the derived photometry.

Table 3 provides the results of the photometry for a total of 1,068,803 sources \footnote{The table is provided in its entirety in electronic form only}. Column 1 gives the field number, column 2 the CCD number within the specified field, column 3  lists the unique ID of the source, columns 4 and 5 give the J2000 coordinates for each star, and columns 6-7 and 8-9 provide the magnitudes and magnitude errors (including zero point error) for the $B$ and $I$ filters, respectively.   

\subsection{Completeness Evaluation}

Extensive artificial star experiments were performed in order to estimate the level of completeness 
of the data. We used the standard artificial star package in DAOPHOT, ADDSTAR. The PSF model adopted for the artificial stars was identical to the one derived for the observed stars in each CCD. In order to minimize the source confusion the number of the artificial stars added per CCD was limited to 8\% of the total number of sources detected in the particular area. The artificial stellar magnitudes were equally distributed within the instrumental magnitude range of the observed stars. The artificial stars were randomly placed on the images, which were subsequently analyzed in exactly the same way as described in Section 2.

An artificial star was considered as recovered by the detection and photometric reduction process, if its final position lay within 0.7 pixels from the original one, where 0.7 pixels are more than twice the positional scatter of the astrometric solution. The percentage of recovered stars defines the completeness, per magnitude bin. The same procedure was repeated 10 times in order to increase the statistical sample and estimate the uncertainty in the completeness values derived. 

The 50\% completeness levels and the corresponding uncertainties for all fields and filters are shown in the last column of Table 2. 
The range of the 50\% completeness level is from 22.3 mag (in $B$, for field SMC3) to 20.4 mag (in $I$, for field SMC6).   It is noted that the  $I$ data are shallower than the $B$ data in all fields, and therefore the 50\% completeness levels occur at brighter magnitudes. 

The left panel of Fig.~\ref{fig3} shows the  50\% completeness level for the 8 different CCDs in the mosaics of the four fields observed, for the $I$ filter as an example. The  right panel shows the dependence of the  50\% completeness level on the projected surface stellar density.  No significant trends are observed in either plot, although there are small variations that can generally be accounted for by the variability in the sky background.

\begin{figure*}
	\includegraphics[height=11cm,width=7cm]{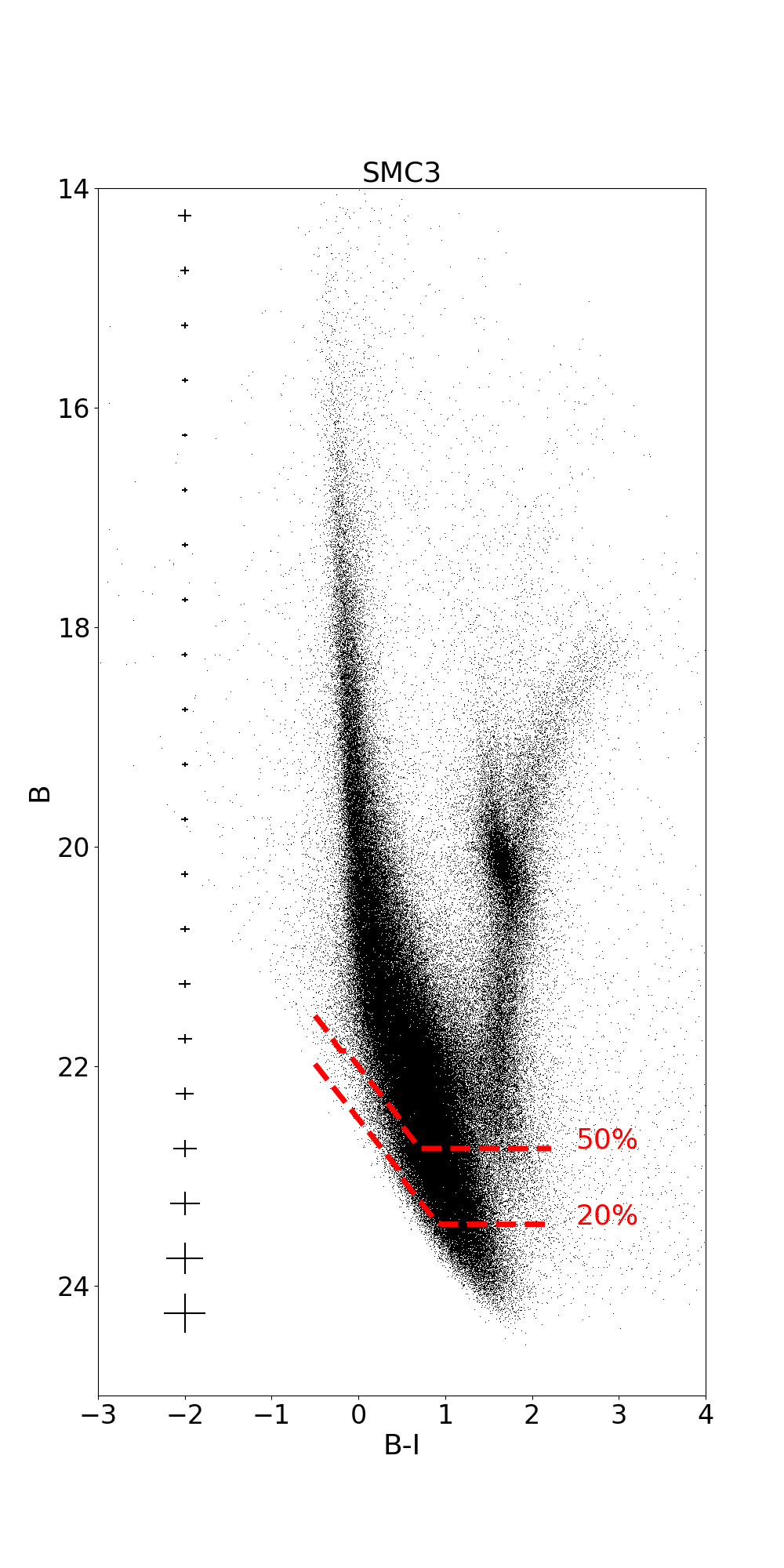}
	\includegraphics[height=11cm,width=7cm]{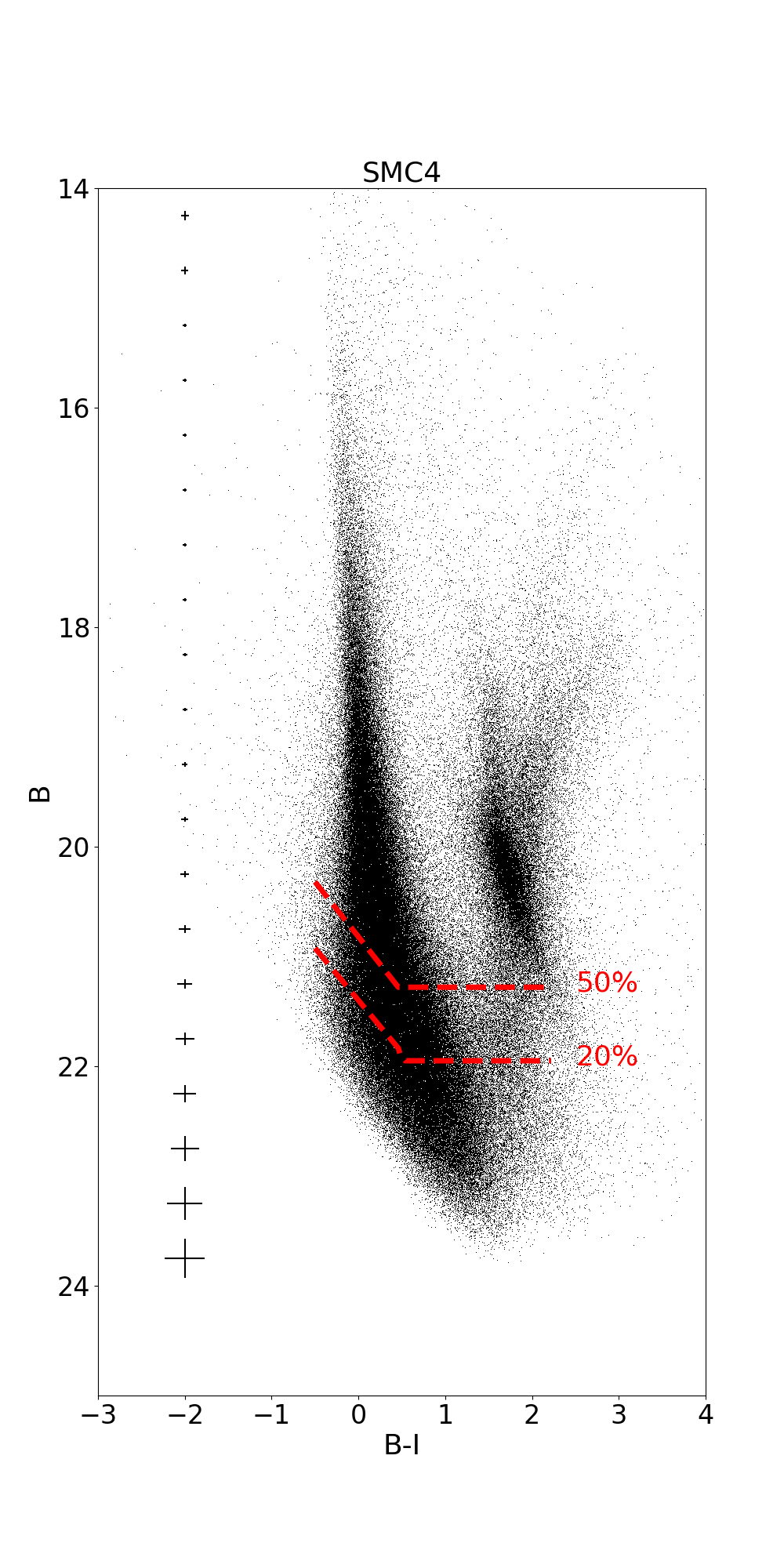} 
	\includegraphics[height=11cm,width=7cm]{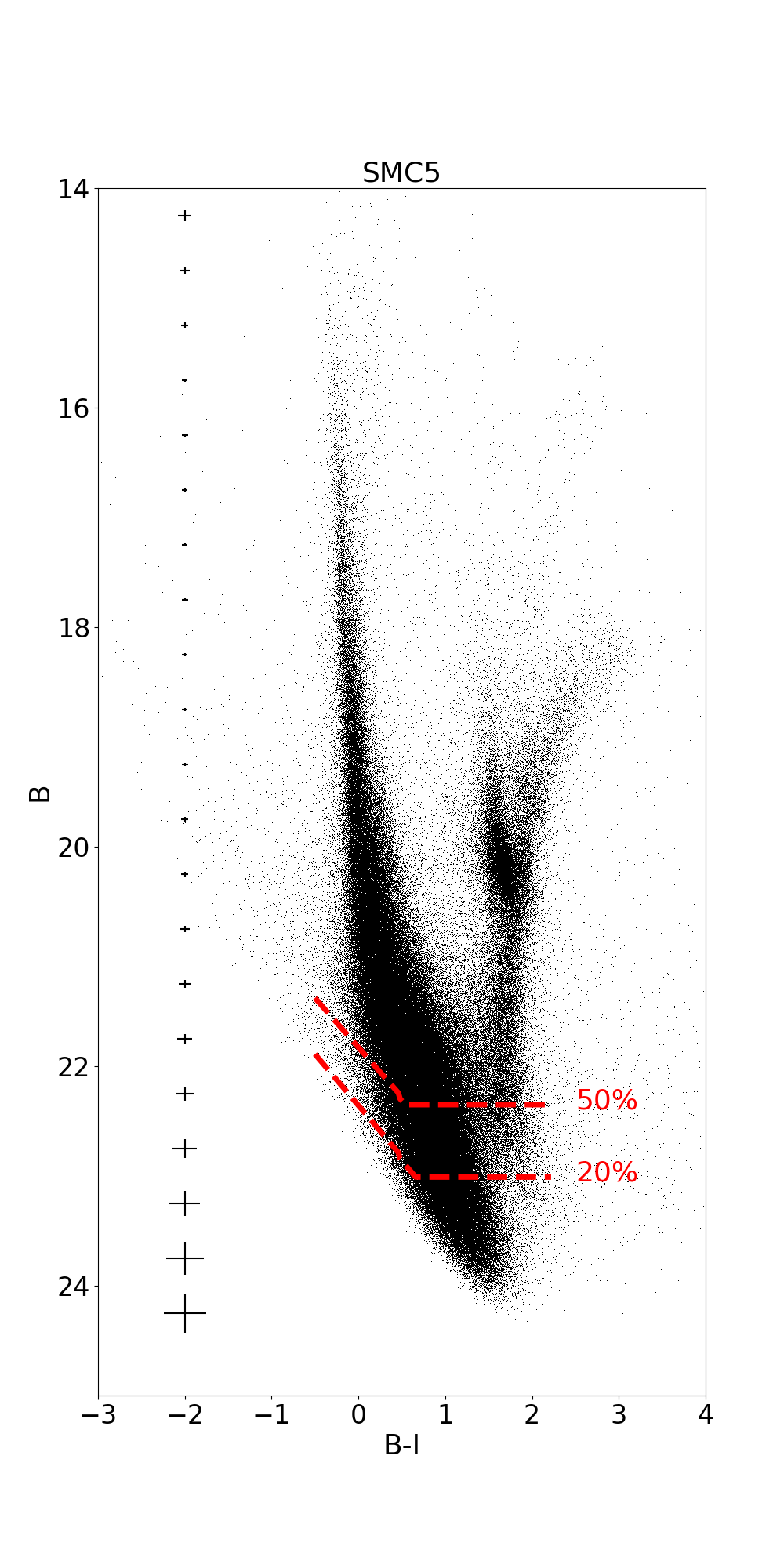}
	\includegraphics[height=11cm,width=7cm]{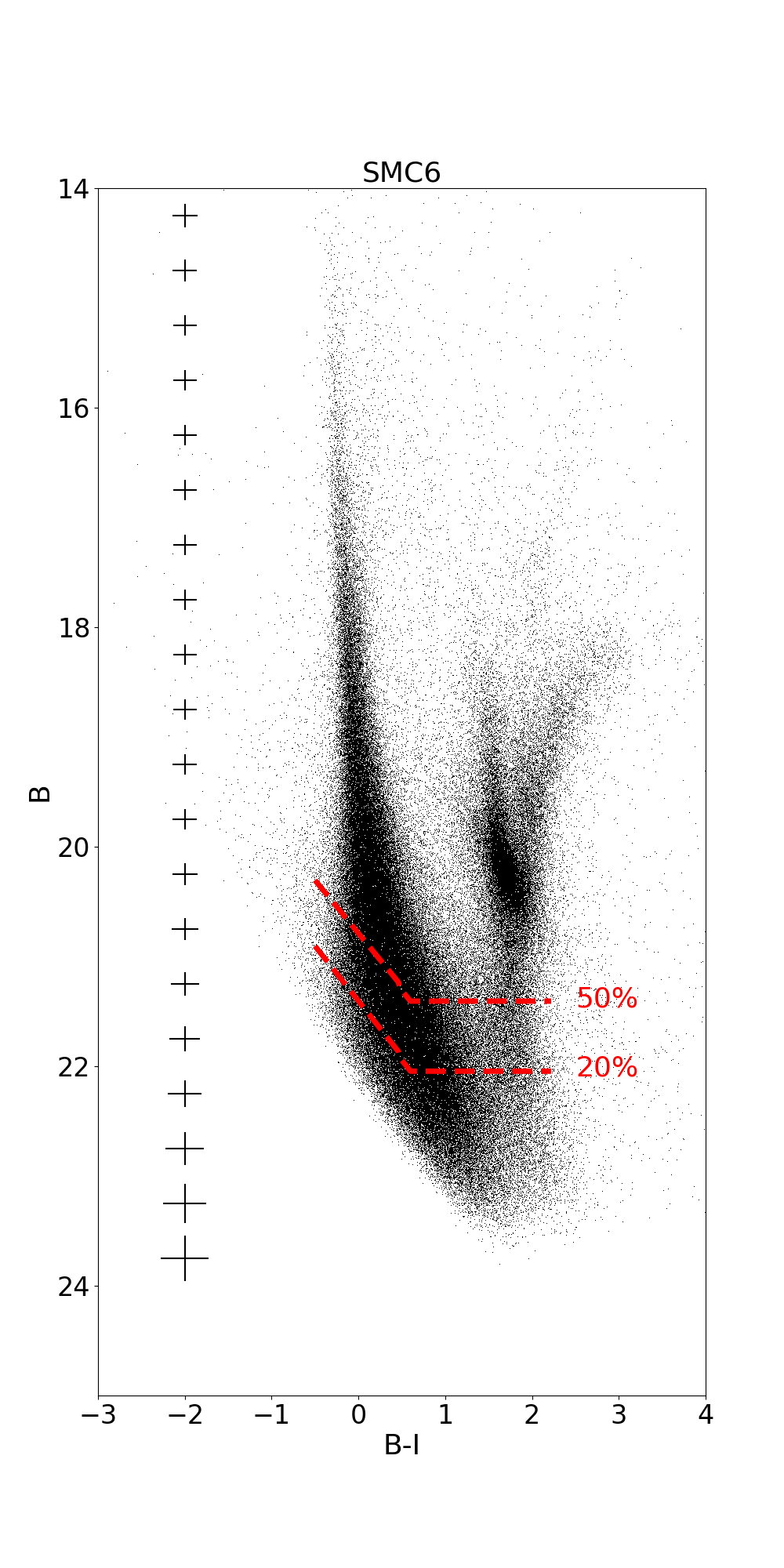}
	\caption{$B$ versus $B-I$ colour-magnitude diagrams in the four studied fields SMC3 (upper left panel), SMC4 (upper right panel), SMC5 (lower left panel), and SMC6 (lower right). The photometric uncertainties are denoted with the error bars shown on the left side of the CMDs. The red dashed lines indicate the 50\% and 20\% completeness levels for each field.}
	\label{fig4}
\end{figure*}

\twocolumn[
  \begin{@twocolumnfalse} 
  \maketitle{Table 2: Photometric analysis statistics.}
	\begin{center}
			\begin{tabular}{ c c c c c c c c c c }
				\hline
				Field & Filter & Zero Point range & N$_{tot}$ & N$_{\sigma\leq0.2}$& 50\% Completeness level\\
				SMC3 & B & $6.051\pm 0.009$ - $6.391\pm 0.013$ & 411074 & 365703 & $22.74\pm 0.10$& \\
				SMC3 & I & $4.636\pm 0.010$ - $4.833\pm 0.009$ & 391140 & 331254 & $21.91\pm 0.11$& \\
				SMC4 & B & $4.813\pm 0.016$ - $5.112\pm 0.010$ & 338836 & 317232 & $21.26\pm 0.06$& \\
				SMC4 & I & $4.324\pm 0.007$ - $4.547\pm 0.010$ & 415748 & 349697 & $20.79\pm 0.07$& \\
				SMC5 & B & $5.506\pm 0.010$ - $5.743\pm 0.009$ & 417994 & 369608 & $22.39\pm 0.15$& \\
				SMC5 & I & $4.783\pm 0.016$ - $5.000\pm 0.013$ & 467574 & 377979 & $21.82\pm 0.10$& \\
				SMC6 & B & $5.172\pm 0.007$ - $5.314\pm 0.011$ & 327808 & 299546 & $21.41\pm 0.06$& \\
				SMC6 & I & $4.043\pm 0.006$ - $4.220\pm 0.013$ & 287903 & 247182 & $20.81\pm 0.04$& \\				
				\hline
			\end{tabular}
	\end{center}
  \end{@twocolumnfalse}
]

\section{Results and Discussion}

\subsection{Field-scale Colour Magnitude Diagrams}

\begin{figure}
	\centering
	\includegraphics[height=7cm,width=8cm]{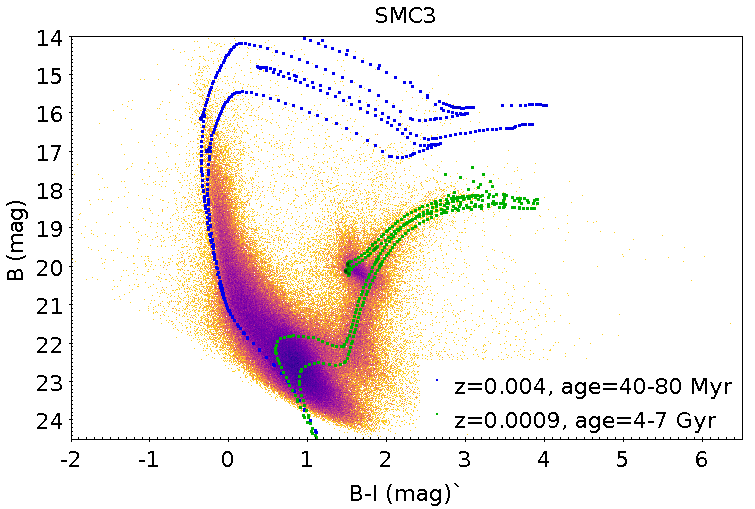}
\caption{CMD of the field SMC3 with PARSEC isochrones of metallicity z=0.0009 for older populations (green lines) and z=0.004 for the younger populations (blue lines). The old-population isochrones are 4 and 7 Gyr, while for the young populations we show the isochrones are ages 40 and 80 Myr. }
\label{fig5}
\end{figure}
We constructed  $B$ versus $B-I$ colour-magnitude diagrams (CMDs) for the four fields (Fig.~\ref{fig4}). Each CMD corresponds to a 0.152 square-degree area. The error bars shown on the left side of the CMDs denote representative photometric uncertainties, which include both instrumental errors and the uncertainty in the zero-point correction. The differences in limiting magnitude and completeness between fields SMC3 and SMC5 on the one hand and SMC4 and SMC6 on the other, which have been discussed in Sections 2 and 3, are also apparent on the CMDs.

The CMDs of Fig.~\ref{fig4} show a well defined main sequence (MS) reaching up to the saturation magnitude of $B\simeq14$ mag, indicating the presence of young populations in these fields. The subgiant (SG) and red giant (RGB) branches are clearly defined. The red clump (RC) is also evident, as well as its  vertical extension to brighter magnitudes caused by the presence of younger stellar populations in the  fields studied. It is noted that in the case of field SMC4, the RC appears to be strongly elongated (and strongly inclined, essentially along the reddening vector) towards redder colours. This is caused by differential interstellar reddening, which is particularly severe in this field (as also mentioned in Section 3). 
The main sequence turnoff region corresponding to intermediate age and older populations is also clearly delineated, especially in fields SMC3 and SMC5.  It is evident, therefore, that stellar populations of different generations are present in the SMC Bar. This is better demonstrated in Fig.~\ref{fig5}  where indicative PARSEC isochrones \citep[PAdova and TRieste Stellar Evolution Code][]{Bressan2012,Marigo2017} of different ages are overlayed on the CMD of SMC3, as an example. We have assumed here a distance modulus of 18.96 mag \citep{Scowcroft2016}, 
an interstellar reddening value of   
$E(B-I)=0.08$mag (with a dispersion of 0.04mag) as derived from the \citealt{Haschke2011} extinction map for this region. These authors have adopted a Milky-Way like dust extinction curve (for which $R_B\approx1.83 $) to derive the interstellar absorption $A_B$. Recently, \citealt{MericaJones2017} found a larger value of $R_{B475}=2.65\pm0.11$ in  the SMC Bar. Assuming this higher value,  interstellar absorption towards field 3 is estimated to be $A_B\simeq0.21$mag, for $E(B-I)=0.08$mag.

The isochrones shown have ages of 4 and 7 Gyr (green lines), and 40Myr and 90 Myr (blue lines)\footnote{The isochrones shown do not change significantly when using the MIST \citep{Dotter2016}, or the Dartmouth models \citep{Dotter2008}}. Although these isochrones are only indicative, we have assumed a lower metallicity of Z=0.0009 for the older isochrones and a higher metallicity of  Z=0.004 for the two younger ones. These assumptions are not intended to constitute an accurate representation of the age-metallicity relation of the SMC, however, they are in broad agreement with the age-metallicity relations derived for SMC clusters and field populations (e.g. \citealt{DaCosta1998}, \citealt{Kayser2007}, \citealt{Rubele2018}), assuming Z$_{\odot}$=0.0152 \citep[][]{Bressan2012}. It is clear that the MSTO region is well bracketed by the 4 and 7 Gyr isochrones, suggesting enhanced star formation during this period. 

It is noted that although for the purposes of this analysis we have used an average value for the distance modulus, it is well known that the SMC displays large line-of sight depth which needs to be taken into account when interpreting a CMD in terms of populations of different ages. For example in Field 5, using RR-Lyrae variables from  \cite{Kapakos2012}, we find a dispersion of about 7 kpc, which is consistent with the values derived for Cepheids in this region by \citet{Subramanian2015}. Such a range in distance would  correspond to an age spread of about $\pm$0.5Gyr (for the older isochrones shown here).

\twocolumn[
  \begin{@twocolumnfalse} 
  \maketitle{Table 3: Catalogue of detected sources (the full catalogue is available online). }
	\begin{center}
			\begin{tabular}{c c c c c c c c c c c}
				\hline
Field & CCD & Source-ID & RA (J2000) & DEC (J2000) & Bmag & e\_Bmag & Imag & e\_Imag \\
3 & 1 & 1 & 13.81901 & -72.11543 & 17.855 & 0.013 & 18.022 & 0.015 \\
3 & 1 & 2 & 13.89466 & -72.11497 & 17.274 & 0.013 & 17.541 & 0.014 \\
3 & 1 & 3 & 13.72003 & -72.15316 & 16.807 & 0.013 & 17.000 & 0.019 \\
3 & 2 & 19356 & 13.95725 & -72.27774 & 19.050 & 0.013 & 19.016 & 0.017 \\
3 & 2 & 19357 & 14.01589 & -72.27467 & 22.627 & 0.05  & 21.48  & 0.10 \\
4 & 1 & 283237& 12.35343 & -73.14637 & 17.966 & 0.012 & 16.631 & 0.016 \\
4 & 1 & 283238& 12.37794 & -73.14379 & 19.292 & 0.013 & 19.213 & 0.04 \\
4 & 1 & 283239& 12.29457 & -73.14603 & 22.31  & 0.05  & 22.02  & 0.17 \\
				\hline
			\end{tabular}
	\end{center}
  \end{@twocolumnfalse}
]

\subsection{Main Sequence Luminosity Functions}

\begin{figure}
	\centering
    \includegraphics[height=4.5cm,width=8cm]{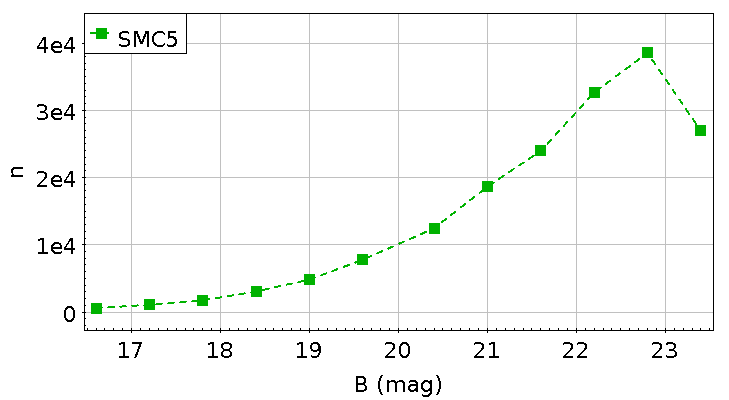}
    \includegraphics[height=4.5cm,width=8cm]{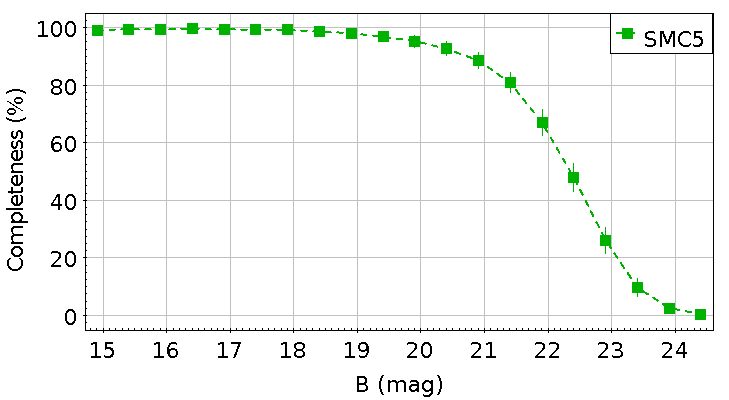}
    \includegraphics[height=4.5cm,width=8cm]{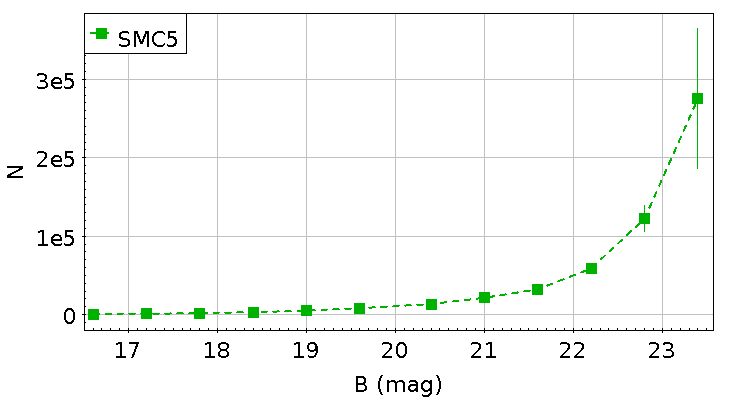}
 \caption{{\it Upper panel}: Luminosity Function of Main Sequence stars in field SMC5; {\it Middle panel}: Variation of completeness as a function of  $B$  magnitude; {\it Lower panel}: Main sequence completeness-corrected Luminosity Function.}
\label{fig6}
\end{figure}
The Luminosity Function (LF) of the MS stars encodes information on the ages of the stellar populations present. Combining the LF of MS stars and the completeness of our data, we derived the completeness corrected LF (CCLF).

Fig.~\ref{fig6} presents the LF of MS stars (upper panel), the completeness curve (middle panel), and the derived CCLF (lower panel), for field SMC5 as an example.

\begin{figure*}
	\includegraphics[height=5cm,width=8cm]{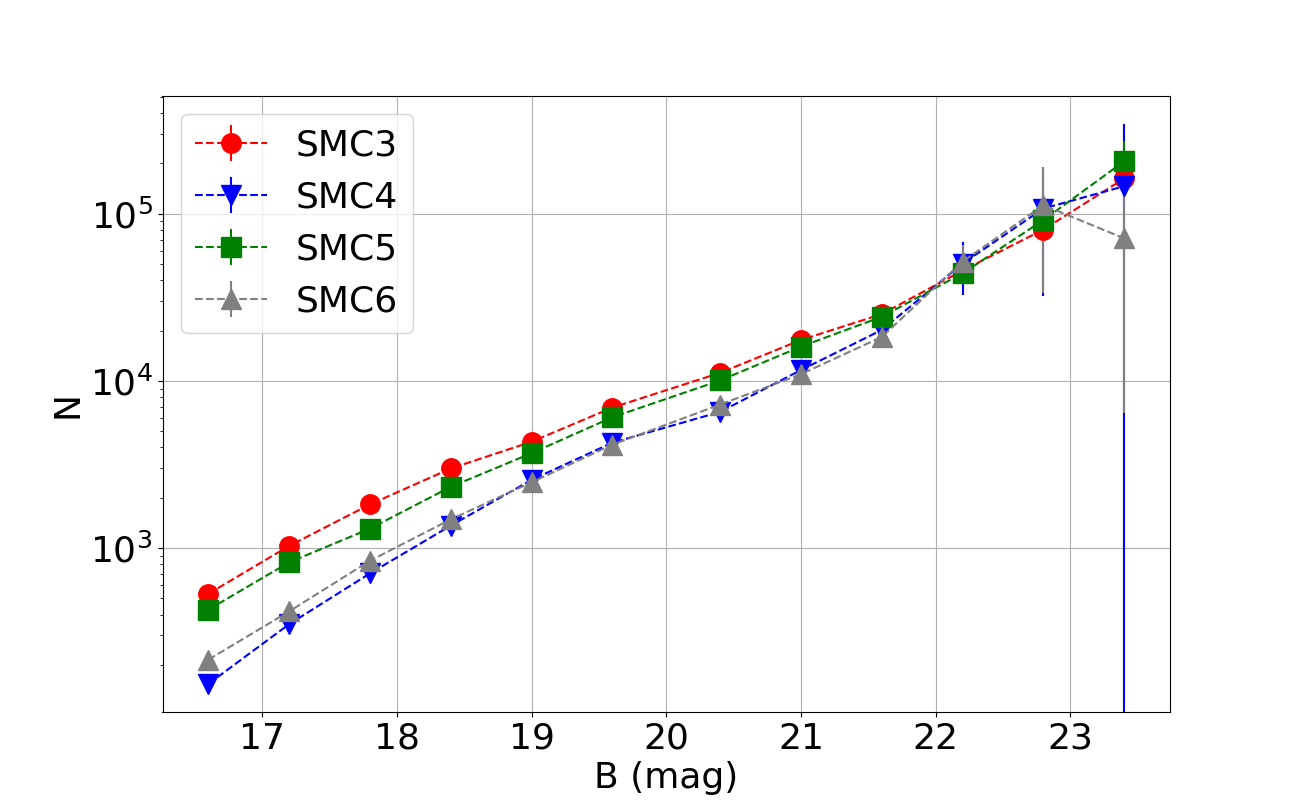}
	\includegraphics[height=5cm,width=8cm]{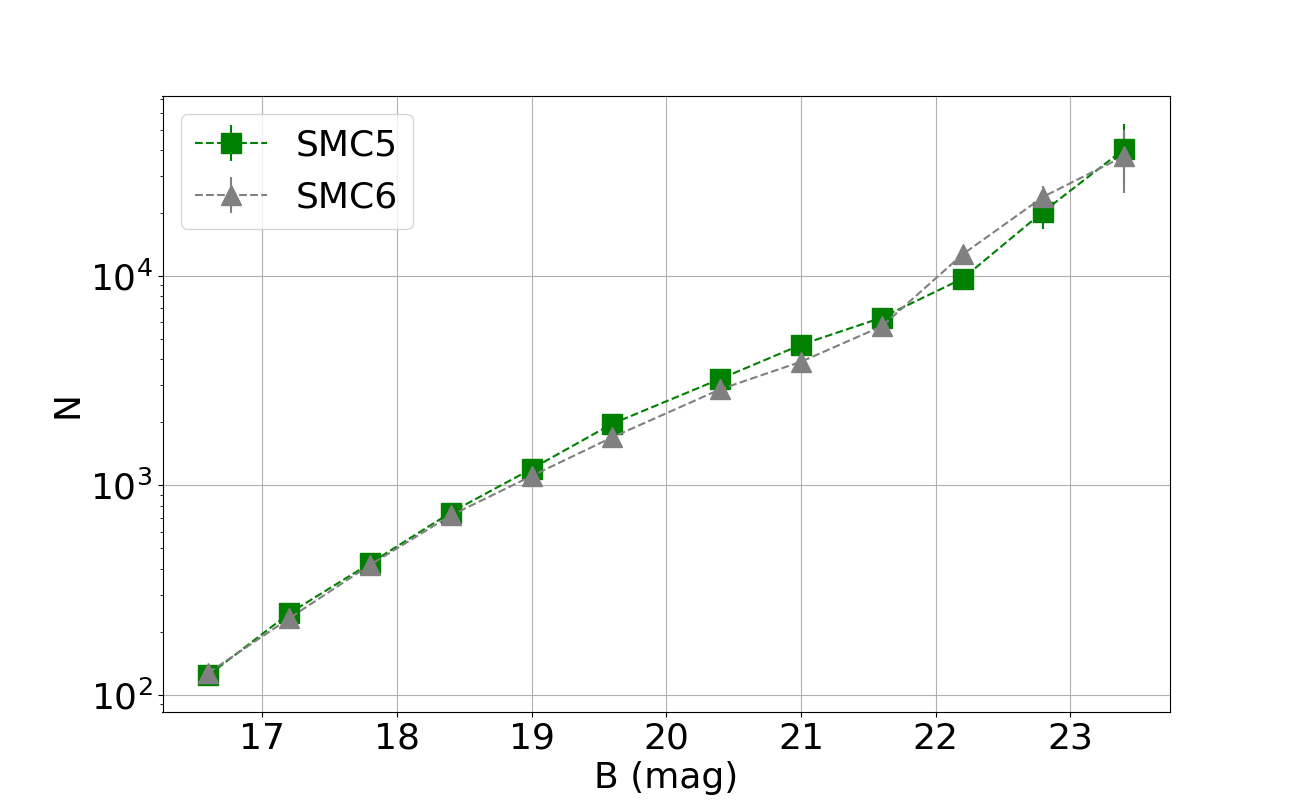}
	\caption{{\it Left panel}:CCLFs of the four studied fields SMC3, SMC4, SMC5 and SMC6. {\it Right panel}: Comparison between of the CCLFs of the common area of the fields SMC5, SMC6.}
	\label{fig7}
 \end{figure*}
The MS CCLF can be used for a rough comparison of the mixture of stellar populations of different ages present in the different fields. Fig.~\ref{fig7} (left panel) shows the MS CCLFs for the four studied fields. The CCLFs have been normalized to that include older stars, i.e., for magnitudes fainter than 21.6 mag in $B$. This normalisation assumes that older populations are more evenly distributed than younger ones. It is immediately apparent that fields SMC3 and SMC5 have a higher contribution from younger populations compared to fields SMC4 and SMC6. This difference was confirmed (at a significance level of 0.05) with the application of a two-sample Kolmogorov-Smirnov test to the cumulative luminosity functions.  It must also be noted that in the overlap region between fields SMC5 and SMC6, the derived CCLFs are identical within the errors (right panel of Fig.~\ref{fig7}). Therefore, the difference between the global CCLF of fields SMC5 and SMC6 cannot be accounted for by uncertainties in completeness and normalisation.  It is noted that fields SMC3 and SMC5 are located to the North-Northeast of the Bar, while fields SMC4 and SMC6 are located towards the South-SouthWest. Younger populations therefore, seem  to be more abundant in the North-NorthEasern regions of the Bar \citep[see also e.g.][]{Rubele2018}.

This qualitative analysis confirms that the history of star formation is not uniform along the SMC Bar. Therefore, the Bar may not be considered as a unique entity, at least in terms of stellar populations

\subsection{Star Formation History}

 We used the method of \citet{Dohm-Palmer1997} to estimate the SFH (i.e. SFR as a function of look-back time) in the four fields studied here. The method is applied to MS stars and it is useful for studying the relatively recent SFH, as the time resolution and accuracy degrade with increasing look-back time.  The same method was recently applied by \citet{Spetsieri2018} to probe the SFH of massive stars in a Virgo cluster galaxy. To derive the SFRs, we follow exactly the same procedures and assumptions described in \citet{Dohm-Palmer1997}, on completeness-corrected MS stars. We binned the MS into discrete regions, assuming that the SFR and stellar mass are approximately constant within each bin.  The colour-magnitude bins have a constant age range. We adopted a \citet{Salpeter1955} initial mass function (IMF), which has a power law index of $\Gamma$=-1.35 and it is not significantly different from the often used \citet{Kroupa2001} or \citet{Chabrier2003} IMFs, for stellar masses down to $\simeq0.8M_{\odot}$.

The application of  the method of \citet{Dohm-Palmer1997} requires knowledge of the relation between $B$ magnitude and stellar mass along the MS, as well as the relation of  $B$ with the MSTO age. Both relations were derived using the PASREC isochrones. Along with the SFR values for different look-back times, we estimated the corresponding uncertainties, taking into account the statistical uncertainty in stellar number counts and in the completeness derivation. 

The Dohm-Palmer procedure implicitly assumes that the calculated SFR in the oldest bins depends on the SFR of the youngest bins. This may lead to an over-subtraction of stars from fainter colour-magnitude bins, rendering negative values for the derived SFRs. Such negative values for the SFRs are observed in fields SMC4 and SMC6, for colour-magnitude bins with stars older than 8 Gyr. Generally, the low number of counts in the youngest bins contribute to the uncertainties in the calculated SFR in all bins. An additional caveat that needs to be mentioned is related to the saturation limit of our photometry at $B\simeq14$mag, which leads to an overestimation of the SFR in the first MS bin considered. 

\begin{figure*}
	\centering
	\includegraphics[height=5cm,width=8cm]{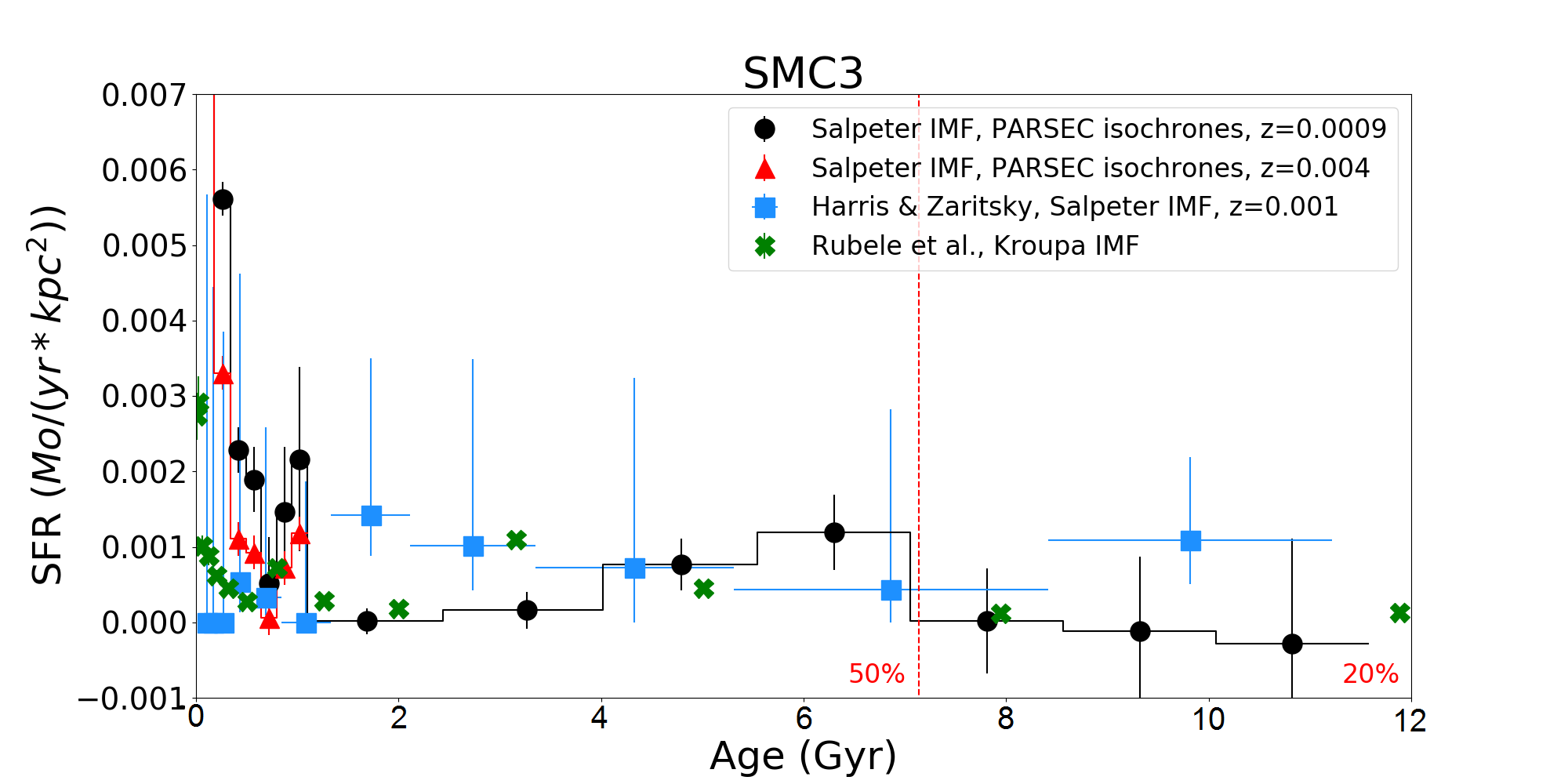}
	\includegraphics[height=5cm,width=8cm]{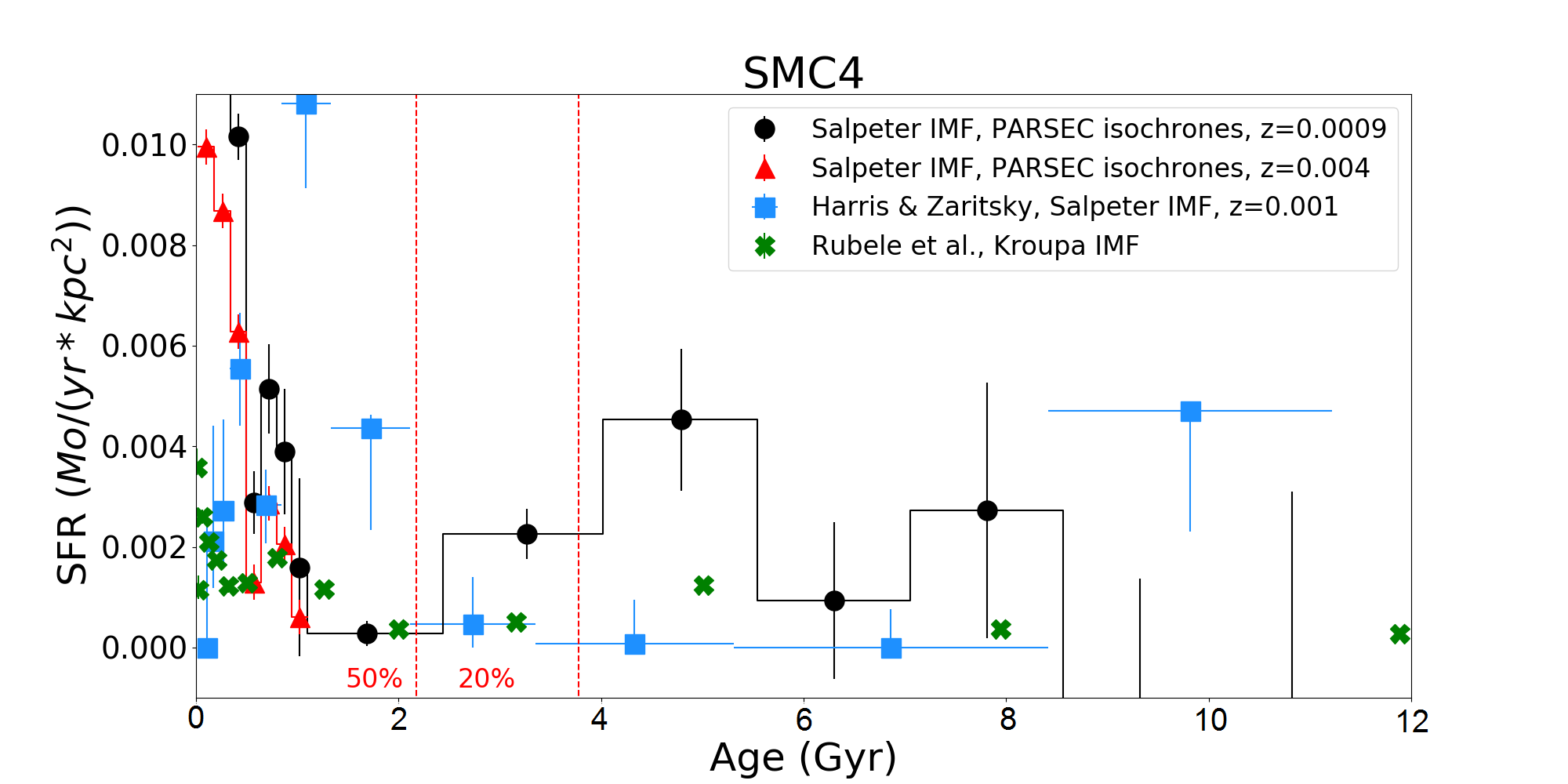}
	\includegraphics[height=5cm,width=8cm]{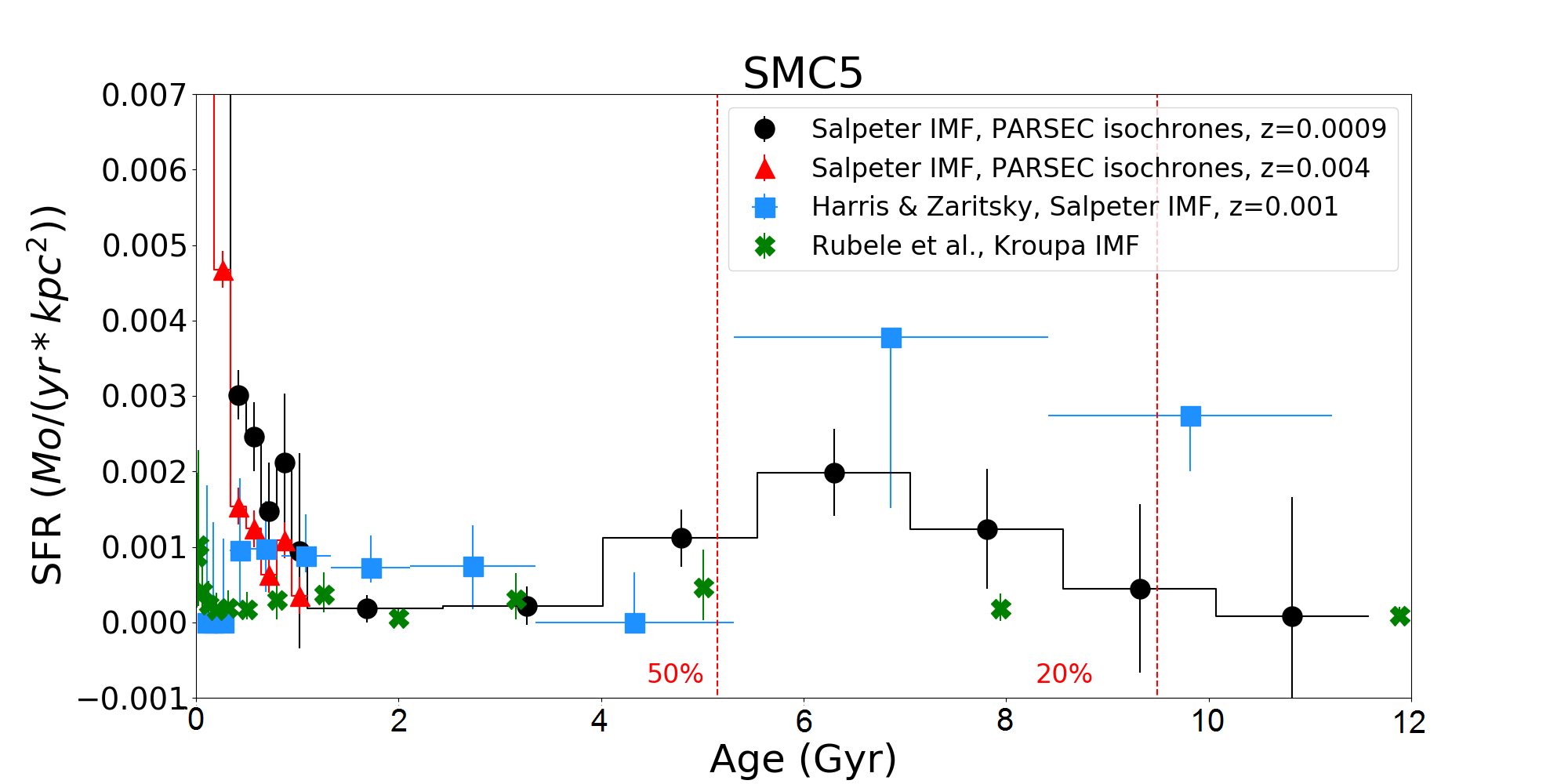}
	\includegraphics[height=5cm,width=8cm]{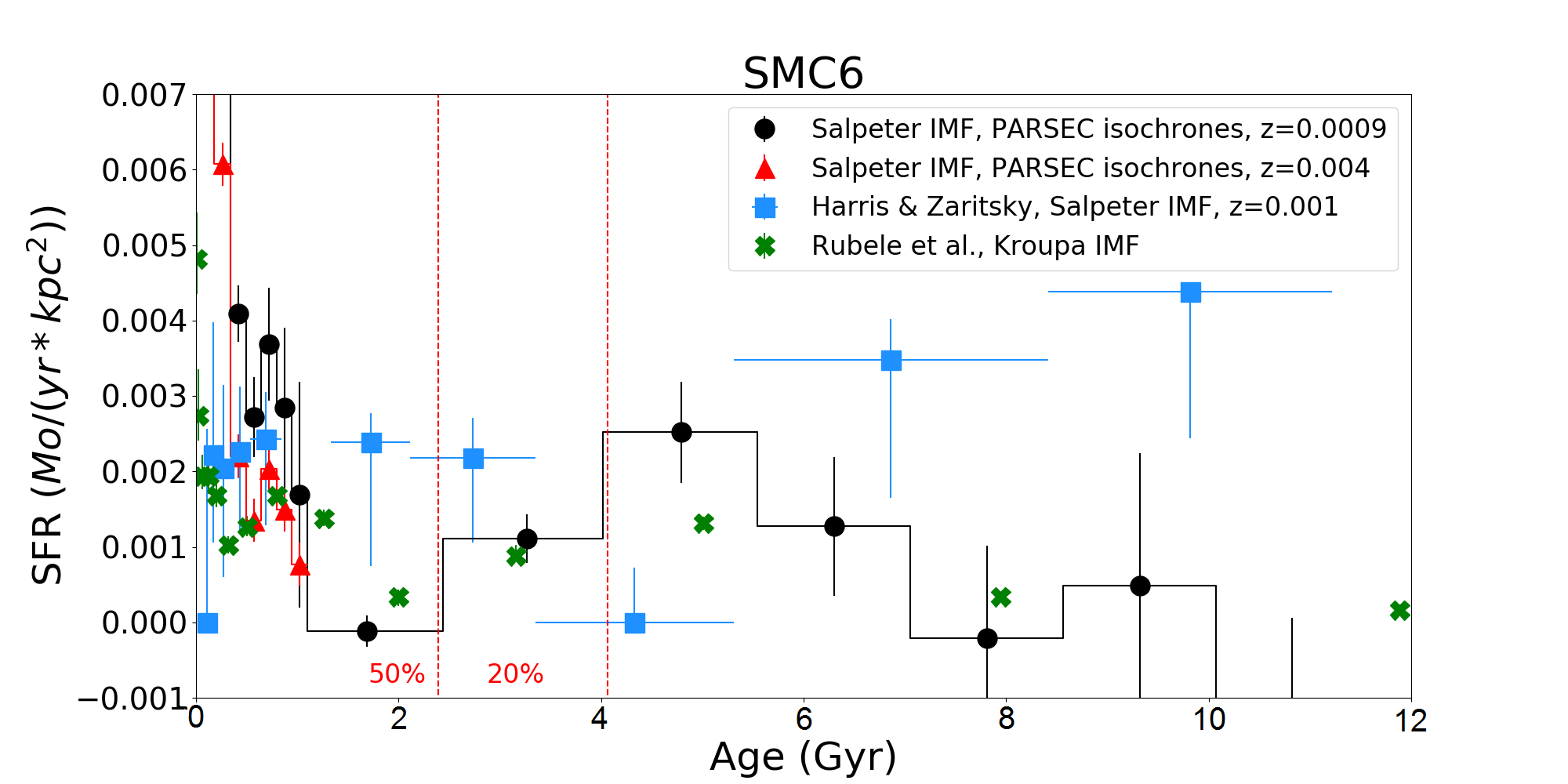}
\caption{SFRs of the four fields, derived from the completeness-corrected LF of MS stars. We use PARSEC stellar evolution tracks adopting a metallicity of Z=0.0009 (black line) and for Z=0.004 (red line). Orange points correspond to the results of \citet{Harris2004} for the same fields as our own, and green points correspond to the results of \citet{Rubele2018}.} 
\label{fig10}
\end{figure*}

Fig.~\ref{fig10} shows the resulting SFH in the four fields, assuming two values for the metallicity, Z=0.004 (red triangles) which is more appropriate for ages below 1Gyr and Z=0.0009 which better matches the metallicity of  older populations (black circles). The choice of these two metallicities has been discussed in section 4.1. We  indicate on the diagrams the location of the 50\% and 20\% completeness limits with vertical dotted lines.  
We also show for comparison the SFH for the same regions from \citet{Harris2004}, who used a Salpeter IMF and a metallicity of Z=0.001, and from \citet{Rubele2018}  who used a Kroupa IMF and a non-constant metallicity. Despite the errors, it is clear that the SFR for the younger populations derived here are too high compared to these studies. This is mainly due to the caveat mentioned earlier, i.e. the fact that we "start" building our SFH for populations older than about 10 Myr, ignoring younger stars, which are present in the fields, but not included in the CMDs due to saturation. Therefore, Fig.~\ref{fig10} can be used to discuss trends of the SFR with time, but not to derive absolute values of the SFR. 

Despite the differences in completeness, a general conclusion that can be drawn from Fig.~\ref{fig10} is that in all four fields star formation has been more intense recently ($<$1Gyr). From about 1 to about 3 Gyr ago SF  activity seems to have been lower, while it peaked again between 4 and 8 Gyr ago. This is in excellent agreement with the appearance of the MSTO region on the field-scale CMDs discussed in section 4.1  In the two fields with fainter completeness limits (fields 3 and 5) the peak in SFR occurs close to an age of 6-7 Gyr. The SFH beyond the 50\% completeness limit is less reliable, therefore for fields 4 and 6 the displacement to somewhat different ages of the SFR peaks beyond 4 Gyr is probably not real.  Small differences can also be expected from variations in interstellar reddening and distance between the different fields. Field SMC4 (and less so, field SMC6) is on average further away than field SMC3 (and less so, than field SMC5) by at least 5kpc, due to the inclination of the SMC Bar \citep{Scowcroft2016}. In addition, interstellar reddening is higher (and variable) particularly in field SMC4, as previously mentioned. Using a larger average  distance (68 kpc) and average reddening (E(B-I)=0.15 mag) value for SMC4 we recalculated the SFH, which was found to be identical  within the errors to the one shown in Fig.~\ref{fig10}.  Therefore, our adoption of a uniform value for the distance and reddening of all fields does not compromise our results.

\begin{figure*}
	\centering
	\includegraphics[height=5cm,width=8cm]{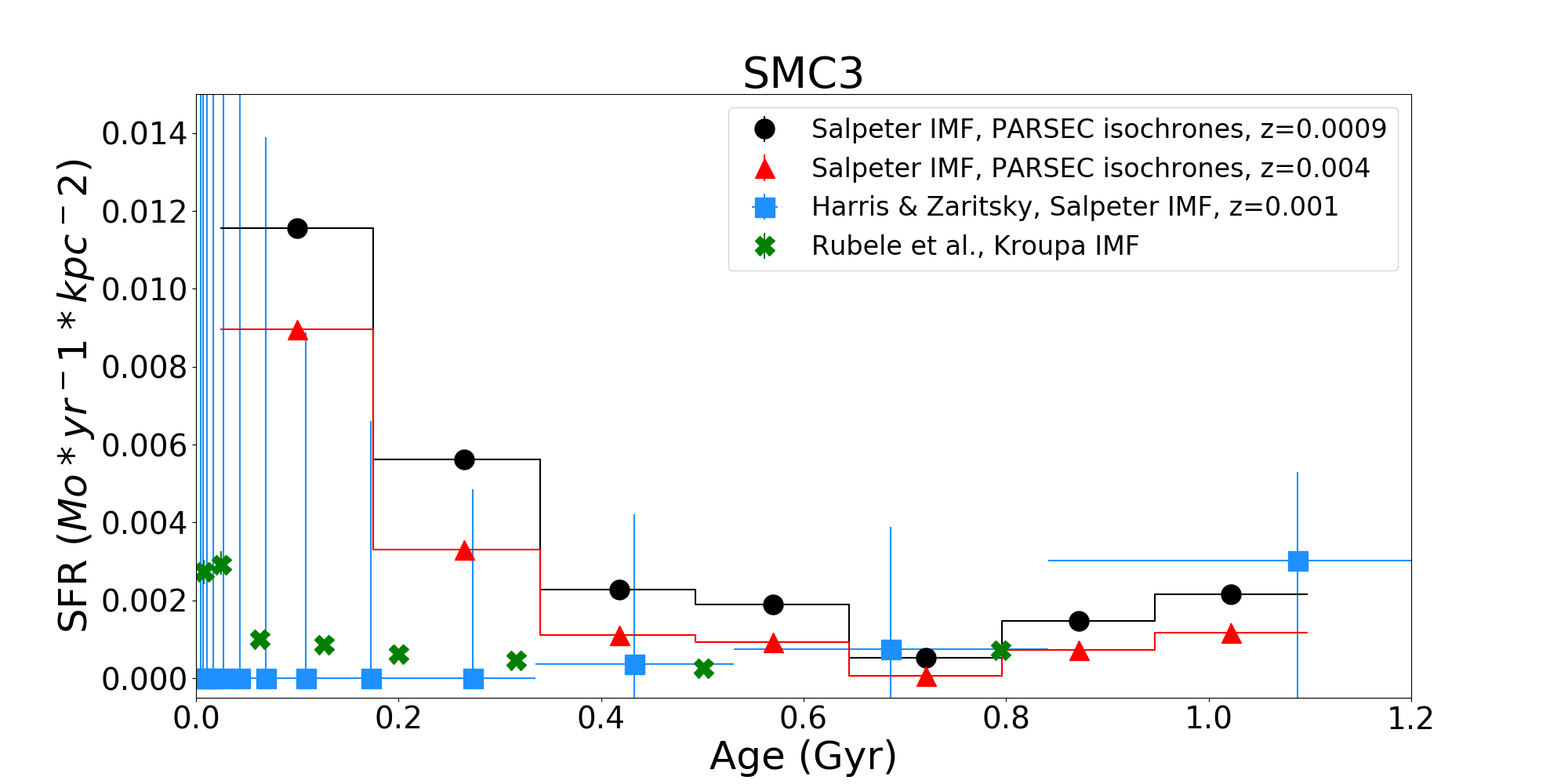}
	\includegraphics[height=5cm,width=8cm]{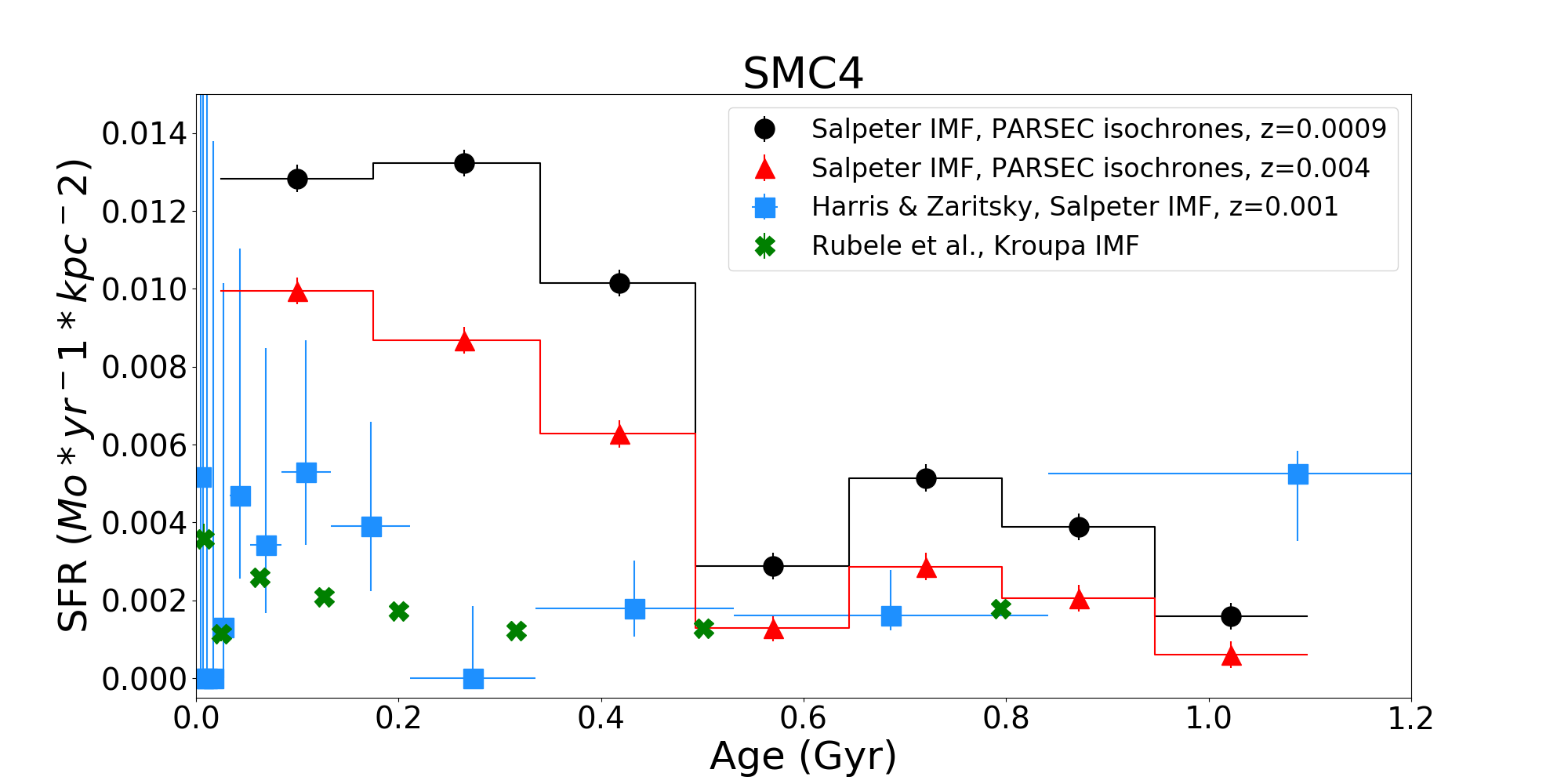}
	\includegraphics[height=5cm,width=8cm]{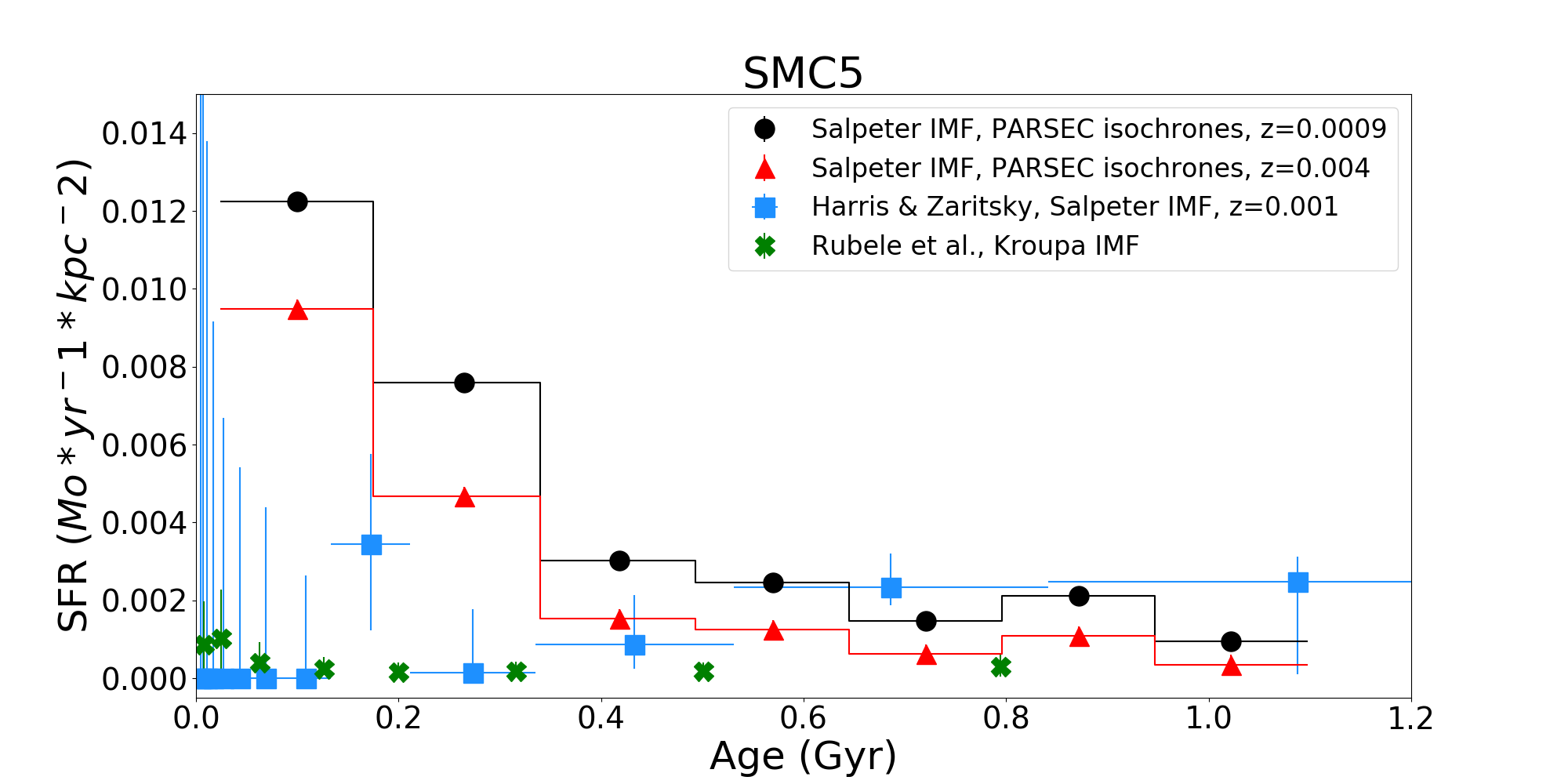}
	\includegraphics[height=5cm,width=8cm]{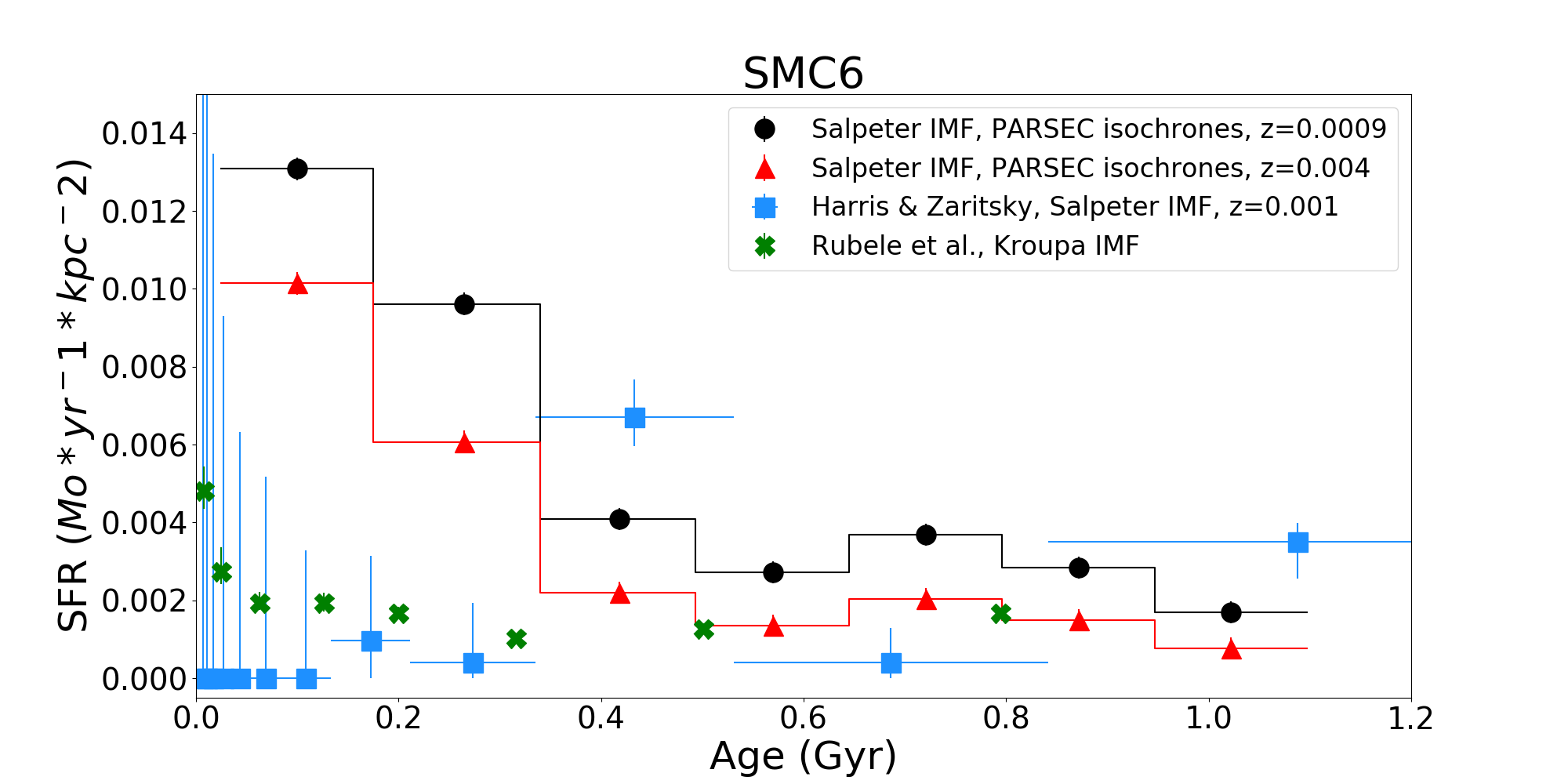}
\caption{SFRs of our four fields zoomed in at younger ages (< 1 Gyr). Orange points correspond to the results of \citet{Harris2004} for the same fields as our own and green points correspond to the results of \citet{Rubele2018}.}
\label{fig11}
\end{figure*}

 In {Fig.~\ref{fig11}}, we show the SFH for stars younger than 1Gyr, in all four fields. These diagrams are zoomed-in versions of the diagrams shown in  Fig.~\ref{fig10}, with the bin width  being smaller. In all fields, the SF activity seems to have been intense over the past $\simeq100$ Myr, decreasing rapidly over a period of few hundred Myr (look-back time). Moreover, in fields 4 and 6, there appears to have been a small enhancement in SF around 700-800 Myr ago.

Generally, many recent studies agree that there are (at least) two major periods of intense SF activity in the SMC main body, one very recently and one around 4-6 Gyr ago \citep[e.g.,][]{Rezaeikh2014, Weisz2013, Cignoni2012, Noel2009}. The fact that SF activity has been very intense over the past 100-200Myr is also corroborated by the burst of cluster formation observed over roughly the same period \citep[e.g.,][]{Bitsakis2018,Nayak2018}. It is interesting to compare our rough estimate of the SFH of the SMC Bar with the results of the full analysis of \citet{Rubele2018} for the same fields. These authors derived the spatially resolved star formation history across the entire main body and Wing of the SMC, using the VISTA survey of the Magellanic Clouds (VMC), in the $YJK_s$ filters and a CMD reconstruction method. As it can be seen in their Figure 11, the SFR decreased from 5Myr to about 300Myr ago (with a small increase around 200 Myr), showed a small enhancement around 800 Myr ago, and then a more significant increase between 3 and 8 Gyr ago. These results are in very good agreement with our findings. On the other hand the results of \citet{Harris2004}, which suffer from much larger uncertainties particularly in the crowded regions studied here, display much poorer agreement with our results.

\subsection{Multiple Main Sequence Turnoffs}

\begin{figure*}
	\centering
	\includegraphics[height=6cm,width=8cm]{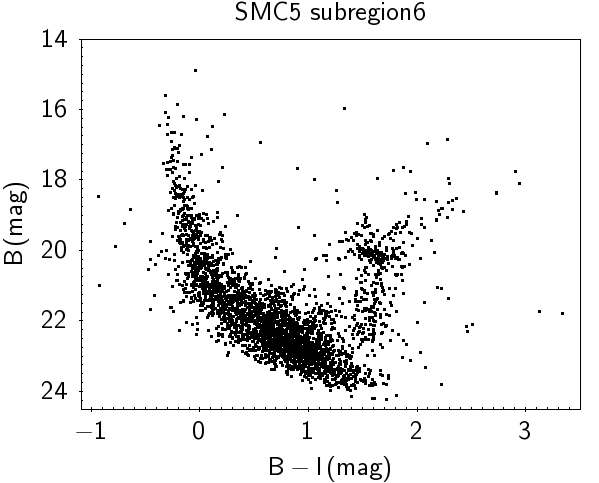}
	\includegraphics[height=6cm,width=8cm]{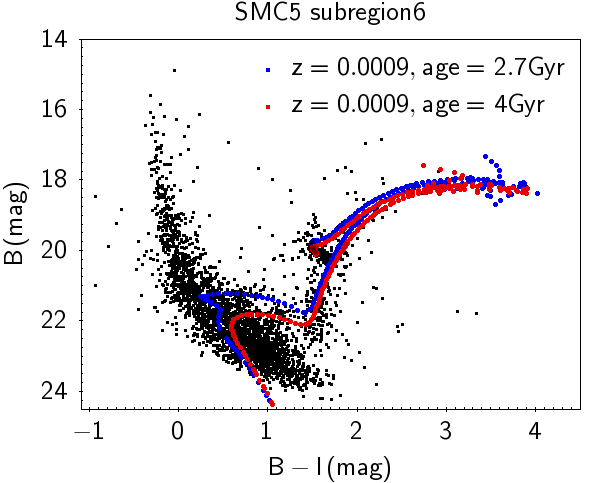}
	\includegraphics[height=6cm,width=8cm]{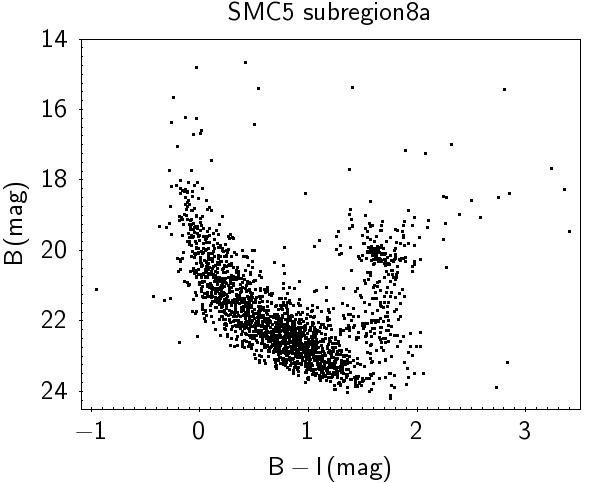}
	\includegraphics[height=6cm,width=8cm]{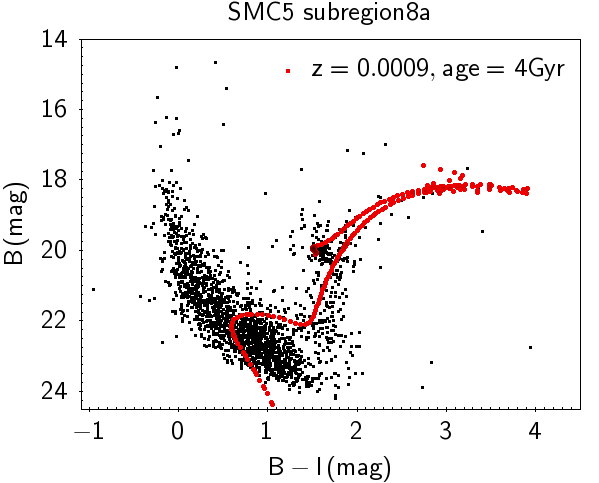}
\caption{ CMDs of subregion 6 (upper panels) and 8a (lower panels) of the field SMC5, with PARSEC isochrones overlaid to probe the detected MSTOs.}
\label{fig8}
\end{figure*}
The global CMDs for the four fields studied may suffer from small but non-negligible residual systematic photometric errors due to variable PSF and completeness levels, differential interstellar absorption and possibly line-of-sight distance variations. All these factors may result in "blurring" specific features,  such as distinct main sequence turnoffs, in CMDs constructed over extended areas.  In order to overcome these limitations, we focused our study on smaller regions with optimal sizes of about 1.1 arcmin in radius.  Regions of this size contain enough stars to allow for detection of stellar evolutionary features in the CMD, and at the same time they are small enough to be less influenced by the previously mentioned systematic effects. Clearly such systematics are likely to manifest themselves when intercomparing separate subregions. 
We inspected the CMDs for $\simeq$1~arcmin subregions in fields SMC3 and SMC5, avoiding star clusters (and their close vicinity) as well as regions severely affected by dust as revealed by the 24$\mu$ emission map (Fig.~\ref{fig1}). We noted that in a few favorable regions (in terms of dust absorption, completeness etc) there are clear indications of distinct MS turnoffs (MSTOs). An example (center of region at RA(J2000) 00:54:42.8 and Dec -72:17:48.8) is shown in the top panels of Fig.~\ref{fig8}, where two distinct MSTOs can be seen.

\begin{figure}
	\centering
	\includegraphics[height=5cm,width=7cm]{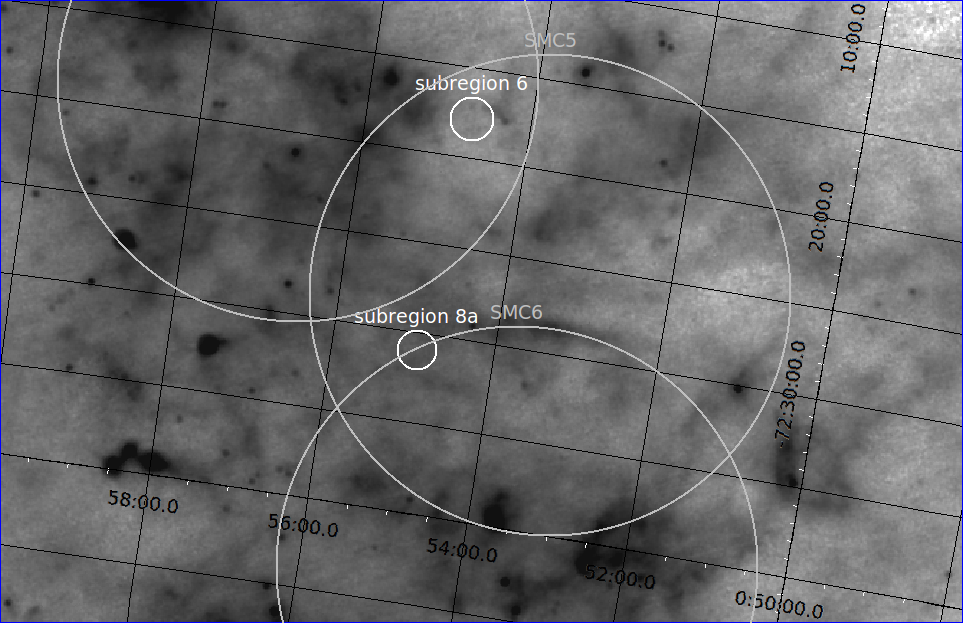}
\caption{ Locations  of the two 1.1 arcmin regions discussed in the text.  }
\label{fig9}
\end{figure}  
We attempted to estimate the average ages of the populations corresponding to these two MSTOs through comparison with PARSEC isochrones with a metallicity of Z=0.0009, which is consistent with the metallicity of the intermediate age and old stellar populations in the SMC \citep[e.g.,][]{DaCosta1998,Piatti2012}. We have assumed a local interstellar reddening value of E(B-I)=0.08 and a distance modulus of 18.96 mag (cf Section 4.3).  On the top right panel of Fig.~\ref{fig8} we have overplotted two indicative isochrones on the CMD, corresponding to ages of 2.7 and 4Gyr, which appear to represent sufficiently well the average populations corresponding to the two MSTOs.  Similar results are obtained using the Dartmouth (\citealt{Dotter2008}) and the MIST (\citealt{Dotter2016},\citealt{Choi2016}) isochrones.

 These age estimates are only indicative. A full analysis of the SFH will be presented in  Strantzalis et al. (2019, in preparation), following the method of Cole (\citealt{Cole2014}, \citealt{Lianou2013}, \citealt{Cignoni2012}, \citealt{Cole2007}). 
 
 In the bottom panel of Fig.~\ref{fig8} we show the CMD in another 1.1 arcmin region (RA(J2000) 00:54:56.0 , Dec(J2000) -72:31:02.0). In this case  there is no clear indication of the younger MSTO at 2.7 Gyr. The  older MSTO corresponding to about 4 Gyr seems to be  prevalent. It is noted that although we have made the same assumptions for the distance modulus and interstellar reddening as in the previous case,  variations may be expected, both in reddening and distance along the line of sight. 
 
 In Fig.~\ref{fig9} we show the location of the two regions of Fig.~\ref{fig8} on the 24$\mu$ map. It is clear that both regions are lying in areas with relatively lower than average dust content. Both regions have a 50\% completeness level at $\simeq$22.5mag.

 The two ages where there seem to be distinct turnoffs lie roughly within the intermediate age peak discussed in section 4.3. \citealt{Cignoni2012} also proposed that there has been a strong  enhancement in SF 4-6 Gyr ago in the Bar. A similar conspicuous peak around the age of 4-5Gyr was found by \citealt{Noel2009}, along with  a less conspicuous one at 1.5-2.5Gyr, but in regions outside the central Bar. Similar enhancements in SF were proposed by \citealt{Rubele2015}. However, in none of these occasions were distinct turnoffs observed. The two turnoffs that we have identified suggest clearly distinct events of SF, separated by about 1.3 Gyr. Observations and theoretical arguments strongly suggest a recent collision between the Clouds in the last 200-300 Myr that formed the Bridge and contributed to the Leading Arm and the trailing Stream \citep{D'Onghia2016}. This time also coincides with strong star and cluster formation activity in the SMC.   Proper motion studies of the Magellanic Clouds \citep{Kallivayalil2013} indicate that the LMC and SMC must have been a bound pair for at least several Gyr (i.e. before their infall to the MW). In this picture, it is possible that the distinct epochs of SF may be linked to close passages between the SMC and the LMC.  In such a scenario, the LMC might also be expected to show similar enhancements in SF as the SMC. However, given the fact that the LMC is about ten times more massive than the SMC, it is very likely that the effect of the close passages on SF would be less marked.  Confirmation or otherwise of this scenario must await for detailed modelling, improved knowledge of the MW potential and of the MC motions.

\section{Summary and Conclusions}
 We have used $B$ and $I$  CCD mosaic images obtained with the 6.5m Magellan Telescope to study the resolved stellar populations in four SMC fields located along the Bar of the SMC, each field tracing different interstellar medium content and projected stellar densities. The main advantage of our study is the combination of  large area coverage with  high angular resolution, which is very important in the dense regions in the SMC Bar. The analysis of the MS luminosity functions showed that the history of star formation is not uniform along the SMC Bar. Therefore, the Bar cannot be considered as a unique entity in terms of stellar populations. There is a clear indication that younger populations have a more significant presence in the north-northeastern fields, as compared to the southern-most field. With the application of  the method of \citet{Dohm-Palmer1997} to the MS stars, we derived the variation of SFR as a function of look-back time in the four fileds studied. This analysis revealed a SFR enhancement within the first 100 Myr, which drops to lower SFRs with look-back time. In the fields SMC4 and SMC6, we detected a second SFR enhancement, which occurs at an age of $\sim$800 Myr. After a period of relatively low SF activity, another significant enhancement occurred  at ages between 4 and 7 Gyr (see Fig.~\ref{fig10}). This period of enhanced SF is also apparent in the MSTO region on the field-scale CMDs.

Finally, we have been able to detect two distinct MSTO corresponding to 2.7Gyr and 4 Gyr  in specific 1.1arcmin-sized  regions where extinction is lower and completeness high.  

Although old stars as exemplified by RR-Lyraes and the old globular cluster NGC121 are present in the SMC, most of the stars seems to have been formed over the past 8 Gyr. There are clear indications that SF has not been continuous over this period and intense SF activity has been followed by periods  of low SF. The connection of this intermittent activity to the orbital history on the Magellanic Cloud system is discussed, but more data on the orbits are necessary before further conclusions can be drawn.

\section*{Acknowledgements}
A.S. acknowledges financial support from State Scholarshpis Foundation, A.Z. acknowledgements funding from the European Research Council under the European Union$\;'$s Seventh Framework Programme (FP/2007-2013)/ERC Grant Agreement n. 617001.  This project has received funding from the European Union's Horizon 2020 research and innovation programme under the Marie Sklodowska-Curie RISE action, grant agreement No 691164 (ASTROSTAT),  V.A. acknowledges financial support from NASA/ADAP grant NNX10AH47G, and the Texas Tech President$\;'$s Office.





\bibliographystyle{mnras}

\bibliography{Bibliography1.bib}

\bsp	
\label{lastpage}
\end{document}